\documentclass{aa}  

\usepackage{graphicx}
\usepackage{txfonts}

\usepackage{color}

\newcommand{\marginnote}[1]{{\boldmath$\color{red}\bigvee$}

\marginpar{\baselineskip3ex{\color{red}#1}}}

\usepackage{natbib}
\bibpunct{(}{)}{;}{a}{}{,} 

\begin{document} 

   \title{The XMM Deep Survey in the CDF-S VII. UV catalogue of the Optical Monitor observations\thanks{Tables 7 and 8 are only available in electronic form
at the CDS via anonymous ftp to cdsarc.u-strasbg.fr (130.79.128.5) or via http://cdsweb.u-strasbg.fr/cgi-bin/qcat?J/A+A/}}

   \subtitle{}

   \author{M. Antonucci
          \inst{1}
          \and
          A. Talavera
          \inst{2}
          \and
          F. Vagnetti
          \inst{1}
          \and
          D. Trevese
          \inst{3}
          \and
          A. Comastri
          \inst{4}
          \and
          M. Paolillo
          \inst{5}
          \and
          P. Ranalli
          \inst{6,4}
          \and
          C. Vignali
          \inst{7,4}
          }

   \institute{Dipartimento di Fisica, Universit\`a di Roma ``Tor Vergata'',
              Via della Ricerca Scientifica 1, 00133 Roma, Italy\\
              \email{marco.antonucci@roma2.infn.it}
         \and
              XMM-Newton Science Operation Centre, ESA, Villafranca del Castillo, Apartado 78, 28691 Villanueva de la Ca\~{n}ada, Spain
         \and
              Dipartimento di Fisica, Universit\`a di Roma ``La Sapienza'',
              Piazzale Aldo Moro 2, 00185 Roma, Italy
         \and
              INAF-Osservatorio Astronomico di Bologna, Via Ranzani 1, 40127 Bologna, Italy
         \and
              Dipartimento di Scienze Fisiche, Universit\`a di Napoli Federico II, Via Cinthia 9, 80126 Napoli, Italy
         \and
             Institute of Astronomy, Astrophysics, Space Applications, and Remote Sensing, National Observatory of Athens, Palaia Penteli, 15236, Athens, Greece  
         \and
              Dipartimento di Fisica e Astronomia, Universit\`a di Bologna, Via Berti Pichat 6/2, 40127 Bologna, Italy
             }

   \date{Received ; accepted }

 
  \abstract
   {The XMM-Newton X-ray observatory has repeatedly observed the Chandra Deep Field-South (CDF-S) in 33 epochs (2001-2010) through the XMM-CDFS Deep Survey \citep{coma11}. During the X-ray observations, XMM-OM \citep{maso01} targeted the central 17$\times$17 arcmin$^2$ region of the X-ray field of view, providing simultaneous optical/UV coverage of the CDF-S. The resulting set of data can be used to build an XMM-OM catalogue of the CDF-S, filling the UV spectral coverage between the optical surveys and GALEX observations.}
   {We present the UV catalogue of the XMM-CDFS Deep Survey. Its main purpose is to provide complementary UV average photometric measurements of known optical/UV sources in the CDF-S, taking advantage of the unique characteristics of the survey (UV and X-ray simultaneous data, time monitoring $\sim$8.5 years, multi-wavelength photometry). The data reduction is also intended to improve the standard source detection on individual observations by cataloguing faint sources by stacking their exposure images.}
   {We re-processed the XMM-OM data of the survey and stacked the exposures from consecutive observations using the standard Science Analysis System (SAS) tools to process the data obtained during single observations. Average measurements of detections with SAS good quality flags from individual observations and from stacked images were joined to compile the catalogue. Sources were validated through the cross-identification within the ESO Imaging survey \citep{arno01} and the COMBO-17 survey \citep{wolf04}.}
   {Photometric data of 1129 CDF-S sources are provided in the catalogue, and optical/UV/X-ray photometric and spectroscopic information from other surveys are also included. The stacking extends the detection limits by $\sim$1 mag in the three UV bands, contributing $\sim$30\% of the catalogued UV sources. The comparison with the available measurements in similar spectral bands confirms the validity of the XMM-OM calibration. The combined COMBO-17/X-ray classification of the ``intermediate'' sources (e.g. optically diluted and/or X-ray absorbed AGN) is also discussed.}
   {}


   \authorrunning{M. Antonucci et al.} \titlerunning{XMMOMCDFS catalogue.}

   \maketitle
%

\section{Introduction}\label{intro}      

The XMM-CDFS Deep Survey was conceived to investigate the X-ray spectral properties of the highly obscured and Compton-thick active galaxies \citep{coma11}. The repeated, long-exposure time observations of the Chandra Deep Field-South (CDF-S) performed by XMM-Newton made it possible to collect the X-ray photons emitted by the heavily obscured sources, notwithstanding the strong absorption due to the obscuring dust. The foremost goal of the survey is to constrain the contribution of such obscured objects to the X-ray unresolved background \citep[e.g.][]{gill07, trei09, akyl12}, and the star formation and accretion processes in the early phases of the AGN evolution \citep[e.g.][]{page04, menc08}. The first results in the analysis of the data were provided by the colour selection \citep{iwas12} and spectroscopic selection \citep{coma11, geor13} of the moderate-to-high redshift samples of highly obscured active galactic nuclei (AGN). Analysis of the iron $K_{\alpha}$ line of the average AGN spectra in the XMM-Newton CDF-S observations was carried out \citep{falo13}. The efficiency of IR power-law spectra, which is the selection criterion for finding obscured X-ray sources, was also studied \citep{cast13}. XMM-Newton point-source catalogues in the hard bands 2-10 keV and 5-10 keV, with the respective number counts, were published by \citet{rana13} and contain 339 (flux limit 6.6$\times$10$^{-16}$ erg/s/cm$^2$) and 137 objects (flux limit 9.5$\times$10$^{-16}$ erg/s/cm$^2$). In particular, the XMM-CDFS represents the deepest XMM-Newton survey currently published in the 5-10 keV band.

During the X-ray observations of the XMM-CDFS Deep Survey, the optical/UV telescope onboard the XMM-Newton observatory \citep[Optical Monitor, XMM-OM,][]{maso01}, co-aligned to the X-ray mirrors, simultaneously targeted the CDF-S with various filter-selection and exposure-time criteria (see Table \ref{tabobsid}). The complementarity of the optical/UV and X-ray observations and the wide time sampling of the survey enables multi-wavelength variability analyses. It is possible, for instance, to study the individual and combined variability of the AGN X-ray/UV ratio by extracting a sample of sources of known redshift \citep[Vagnetti et al., in preparation; see also][]{vagn10, vagn13}. The resulting set of XMM-OM data, however, can also be used to build a catalogue of the CDF-S sources in the optical and UV domains, taking advantage of the wavelength coverage provided by the six filters (three UV and three optical filters, with an overall range 180-590 nm), the full width at half-maximum (FWHM) of the point spread function (PSF) that is smaller than 2 arcsec for each filter, and the 17$\times$17 arcmin$^2$ field of view (see \citet{page12}).

Several optical/near-UV surveys have targeted the CDF-S and its flanking fields, with the main purpose of photometrically and spectroscopically calculating the redshifts of the detectable sources. In the context of the ESO imaging survey \citep[EIS,][]{arno01} project, multi-colour (six broad band filters, U'UBVRI) photometric measurements of about 75\,000 sources have been provided to observe the Extended-CDF-S (E-CDF-S) through the Wide Field Imager \citep[WFI,][]{baad98, baad99} at the MPG/ESO 2.2m telescope on La Silla. By means of the same telescope and instrument, a nearly coincident area of 31.5$\times$30 arcmin$^2$ has been observed in 17 passbands (350-930 nm) within the COMBO-17 survey, obtaining a catalogue of more than 60\,000 objects including astrometry, classification, and photometric redshifts, with around 10\,000 galaxies identified at R$_{Vega}$<24 \citep{wolf04, wolf08}. Another optical and near-IR multi-wavelength survey has been developed by \citet{card10} as part of the Multi-Wavelength Survey by Yale-Chile \citep[MUSYC,][]{gawi06}, combining the Subaru 18-bands imaging (400-900 nm) to 14 other available ground-based and Spitzer images: about 80\,000 galaxies are catalogued at R$\le$27, and about 30\,000 redshifts are computed. The Great Observatories Origins Deep Survey \citep[GOODS,][]{giav04}, instead, covers a 320 arcmin$^2$ region in the CDF-S (GOODS-S) by using the Hubble Space Telescope (HST) Advanced Camera for Surveys \citep[ACS,][]{ford03}, through the filters F435W, F606W, F775W, and F850LP; it catalogued about 29\,600 objects, at m$\le$27. Moreover, an 11 arcmin$^2$ region within the GOODS-S \citep[Hubble Ultra Deep Field, UDF,][]{beck06} has been deeply observed by the ACS in the same bands, reaching a uniform limiting magnitude m$\sim$29 for point sources. \citet{voye09} also presented the U-band observations of the same area, achieved through the HST Wide-Field Planetary Camera 2 (WFPC2)/F300W filter ($\lambda_{max}$$\sim$300 nm) and providing a catalogue of 96 objects. The Arizona CDFS Environment Survey \cite[ACES,][]{coop12} spectroscopically targeted the E-CDF-S by means of the Inamori-Magellan Areal Camera and Spectrograph on the Magellan-Baade telescope (565-920 nm), yielding more than 5000 redshifts for R$\le$24.1 sources; it widens therefore previous sets of spectroscopic redshift estimates of CDF-S sources  \citep[e.g.][]{szok04, lefe05, ravi07, pope09, bale10, silv10}, for a total of about 1300 ACES objects with comparison published redshifts. All the publicly available spectroscopic redshift estimates in the CDF-S have been compiled in the ESO CDF-S Master Catalogue, version 3.0\footnote{http://www.eso.org/sci/activities/garching/projects/goods

/MasterSpectroscopy.html}.

\begin{table*}
\caption{XMM-OM XMM-CDFS Deep Survey}             
\label{tabobsid}      
\centering                          
\begin{tabular}{c c c l l r r r r r r r}
\hline\hline
Group & OBSID & time & RA & Dec. & p.a. & UVW2 & UVM2 & UVW1 &\; U &\; B &\; V\\    
 & & & \multicolumn{3}{c}\hrulefill & \multicolumn{6}{c}\hrulefill \\
 & & & \multicolumn{3}{c} {degrees} & \multicolumn{6}{c} {ks} \\
\hline
   B1 & 0108060401 & 2001-07-27 & 53.1064 & -27.8127 & 58 & 14.4 & 8.0 & & & & \\
   B1 & 0108060501 & 2001-07-27 & 53.1173 & -27.8178 & 58 & 13.1 & 8.2 & 8.2 &\; 4.1 &\; 4.1 & \\ 
   B2 & 0108060601 & 2002-01-13 & 53.1198 & -27.8051 & 237 & 15.0 & 5.0 & 5.0 &\; 4.6 &\; 3.0 & \\ 
   B2 & 0108060701 & 2002-01-14 & 53.1147 & -27.8018 & 237 & 20.0 & 20.0 & 10.0 &\; 5.0 &\; 5.0 & \\ 
   B2 & 0108061801 & 2002-01-16 & 53.1192 & -27.7994 & 237 & 11.9 & 5.0 & 5.0 &\; 2.0 &\; 2.0 & \\ 
   B2 & 0108061901 & 2002-01-17 & 53.1197 & -27.7935 & 237 & 10.0 & 5.0 & 5.0 &\; 2.0 &\; 2.0 & \\ 
   B2 & 0108062101 & 2002-01-20 & 53.1252 & -27.7967 & 237 & 9.1 & 7.6 & 5.0 &\; 2.0 &\; 2.0 & \\ 
   B2 & 0108062301 & 2002-01-23 & 53.1206 & -27.7949 & 237 & 5.0 & 5.0 & 5.0 &\; 2.0 &\; 2.0 & \\ 
   C1 & 0555780101 & 2008-07-05 & 53.1719 & -27.7590 & 60 & & 31.5 & 44.0 &\; 44.0 & & \\ 
   C1 & 0555780201 & 2008-07-07 & 53.1729 & -27.7680 & 60 & & 31.8 & 44.0 &\; 44.0 & & \\ 
   C1 & 0555780301 & 2008-07-09 & 53.1630 & -27.7669 & 60 & & 30.7 & 42.5 &\; 37.0 & & \\ 
   C1 & 0555780401 & 2008-07-11 & 53.1629 & -27.7594 & 60 & & 30.3 & 40.0 &\; 39.0 & & \\ 
   C2 & 0555780501 & 2009-01-06 & 53.1071 & -27.8149 & 240 & 24.0 & 16.0 & 4.0 &\; 4.0 &\; 8.0 &\; 4.0 \\ 
   C2 & 0555780601 & 2009-01-10 & 53.1089 & -27.8220 & 240 & 30.5 & 12.0 & 4.0 &\; 4.0 &\; 8.0 &\; 8.0 \\ 
   C2 & 0555780701 & 2009-01-12 & 53.1069 & -27.8330 & 240 & 34.5 & 12.0 & 4.0 &\; 4.0 &\; 8.0 &\; 4.0 \\ 
   C2 & 0555780801 & 2009-01-16 & 53.0979 & -27.8140 & 240 & 32.0 & 10.0 & 4.0 &\; 4.0 &\; 8.0 &\; 8.0 \\ 
   C2 & 0555780901 & 2009-01-18 & 53.0977 & -27.8229 & 240 & 30.7 & 12.0 & 4.0 &\; 4.0 &\; 8.0 &\; 8.0 \\ 
   C2 & 0555781001 & 2009-01-22 & 53.0977 & -27.8310 & 240 & 34.1 & 12.0 & 3.5 &\; 3.5 &\; 7.5 &\; 7.5 \\ 
   C2 & 0555782301 & 2009-01-24 & 53.0978 & -27.8313 & 240 & 34.0 & 12.0 & 3.5 &\; 3.5 &\; 7.5 &\; 3.5 \\ 
   C3 & 0604960301 & 2009-07-05 & 53.1737 & -27.7774 & 61 & 18.8 & & 50.0 &\; 38.1 & & \\ 
   C3 & 0604960201 & 2009-07-17 & 53.1645 & -27.7672 & 61 & 23.5 & & 35.6 &\; 50.0 & & \\ 
   C3 & 0604960101 & 2009-07-27 & 53.1734 & -27.7670 & 61 & 21.1 & & 50.0 &\; 50.0 & & \\ 
   C3 & 0604960401 & 2009-07-29 & 53.1640 & -27.7766 & 61 & 16.1 & & 35.0 &\; 50.0 & & \\ 
   C4 & 0604961101 & 2010-01-04 & 53.1078 & -27.8060 & 244 & 5.6 & & 50.0 &\; 50.0 & & \\ 
   C4 & 0604961201 & 2010-01-08 & 53.0961 & -27.8060 & 244 & 31.9 & & 25.0 &\; 50.0 & & \\ 
   C4 & 0604960701 & 2010-01-12 & 53.0989 & -27.8156 & 244 & 5.6 & & 45.0 &\; 50.0 & & \\ 
   C4 & 0604960501 & 2010-01-18 & 53.1076 & -27.8151 & 244 & & & &\; 39.9 & & \\ 
   C4 & 0604961301 & 2010-01-19 & 53.1077 & -27.8145 & 244 & & & 13.0 &\; 5.0 & & \\ 
   C4 & 0604960601 & 2010-01-26 & 53.1079 & -27.8232 & 244 & 18.8 & & 50.0 &\; 37.9 & & \\ 
   C4 & 0604960801 & 2010-02-05 & 53.1079 & -27.8336 & 244 & 21.5 & & 40.5 &\; 45.0 & & \\ 
   C4 & 0604960901 & 2010-02-11 & 53.0985 & -27.8338 & 244 & 8.6 & & 45.0 &\; 36.0 &\; &\; \\ 
   C4 & 0604961001 & 2010-02-13 & 53.0980 & -27.8240 & 244 & 21.5 & & 45.0 &\; 36.0 & & \\ 
   C4 & 0604961801 & 2010-02-17 & 53.0983 & -27.8240 & 244 & 22.1 & & 45.0 &\; 45.0 & & \\ 
\hline
\end{tabular}
\tablefoot{The table reports the groups to which the observations belong (Group), the observation identification numbers (OBSID), the dates (time), the pointing sky coordinates (RA and Dec.), the position angle (p.a.), and the actual exposure times for each filter (the last six columns).}
\end{table*}

Looking more specifically at the UV domain, the Galaxy Evolution Explorer \citep[GALEX,][]{mart05} has been surveying the sky in two UV broad bands (far-UV, FUV, 134-179 nm; near-UV, NUV, 177-283 nm), reaching limiting magnitudes of about 20-21 mag (in the FUV and NUV pass-bands, respectively) through the All-Sky Imaging Survey (AIS, $\sim$22\,000 deg$^2$), and about 22.7 mag through the Medium-depth Imaging Survey (MIS, $\sim$2250 deg$^2$), which were published by \citet{bian14a}. The GALEX observations of the CDF-S, instead, mainly refer to the Deep Imaging Survey (DIS, $\sim$80 deg$^2$), whose observation integration time is typically 30 ks: within a $\sim$2830 arcmin$^2$ region, UV photometric measurements of more than 36\,000 sources are collected, at m$\la$26.5 in both bands. The CDF-S GALEX data are available on GalexView\footnote{http://galex.stsci.edu/GalexView/}. We also notice that a $\sim$7 deg$^2$ area centred on the CDF-S is part of the GALEX Time Domain Survey (TDS, $\sim$40 deg$^2$), which has been planned to systematically study the UV variability on timescales of days to years \citep{geza13}: about 250 TDS variable sources are selected in the region including the CDF-S.

Regarding the CDF-S coverage in the other spectral ranges, we limit ourselves to mentioning the deep Chandra surveys in the X-ray band \citep[1, 2, and 4 Ms source catalogues respectively described in][]{giac02, luo08, xue11}. The 4 Ms survey is characterised by a total solid angle coverage of about 465 arcmin$^2$, and its main catalogue includes 740 objects with up to three standard X-ray band measurements (0.5-8 keV, 0.5-2 keV, 2-8 keV). A combined classification of the sources, based on X-ray features and complementary optical spectroscopic information, is also provided: more than 75\% of them are likely AGN. Comparison between the 4 Ms Chandra survey and the XMM-CDFS Deep Survey is described by \citet{rana13}. Notwithstanding the deeper sensitivity of the Chandra surveys, the point-source catalogues by \citet{rana13} include 15 sources not detected by Chandra (5 of which are within the Chandra field of view).

Despite the deep, multi-wavelength sampling of the CDF-S performed in recent years, it appears to be of general interest to develope an XMM-OM catalogue of the XMM-CDFS Deep Survey. As previously mentioned, the XMM-OM characteristics offer a powerful survey capability \citep{kunt08, page12}, particularly in the UV domain, representing an effective compromise between the possibility of targeting extended fields --- as, for instance, in the case of GALEX or the large ground-based telescopes --- and the advantage of a finer spatial sampling with respect to GALEX \citep{page12}. Furthermore, the repeated observations of the field within the XMM-CDFS Deep Survey make it possible to stack the exposures of different observations, in order to improve the source detection and therefore the depth of the survey itself, allowing fainter sources to emerge above the significance threshold. The stacking procedure adopted for adding the exposure images uses the tools of the XMM standard Science Analysis System\footnote{http://xmm.esac.esa.int/sas/} (SAS) to process the data obtained during single XMM-OM observations. The main purpose of the catalogue is to provide complementary UV photometric measurements of known optical/UV sources in the CDF-S; moreover, by comparing the XMM-OM data with the available measures in overlapped, nearly coincident wavelength bands, it is possible to test the XMM-OM calibration.

This paper is structured as follows. The data processing and the construction of the catalogue, which we call XMMOMCDFS, are described in Section \ref{proc}. In Section \ref{ana} we present the statistical description of the catalogue and cross-correlations and comparisons with other surveys in the field and discuss the available classifications of the sources. Section \ref{concl} summarises the results and the future developments in analysis of the XMM-OM CDFS data.

Throughout the paper, we adopt the cosmology H$_0$$=$70 km/s/Mpc, $\Omega_m$$=$0.3, and $\Omega_{\Lambda}$$=$0.7.

%

\section{Observations and data reduction}\label{proc}

\subsection{The XMM-CDFS Deep Survey XMM-OM data}\label{sas}

The XMM-CDFS Deep Survey is composed of six groups of observations, performed at intervals of about six months, with a longer time gap between the first two groups (thereafter B1 and B2, July 2001-January 2002, 8 observations, P.I.: J. Bergeron) and the other ones (C1, C2, C3, and C4, July 2008-February 2010, 25 observations, P.I.: A. Comastri). The XMM-OM data have been obtained adopting different combinations of filters and observational strategies. In two of the groups of observations (B2 and C2), the full XMM-OM field of view is covered simultaneously (``full frame'' configuration), while in the other cases it is targeted by a series of consecutive single exposure frames in which it can be divided (``OM image'' configuration), covering 92\% of its extension; details of the imaging configurations are reported in the XMM-Newton Users Handbook\footnote{Issue 2.12, 2014 (ESA: XMM-Newton SOC), Section 3.5.9.2,

http://xmm.esac.esa.int/external/xmm\_user\_support\\/documentation/uhb/index.html}. Both configurations refer to the ``image mode'': events are spatially recorded, and therefore the coordinates in the bidimensional arrays of data correspond to their detector positions \citep{maso01}. We refer to \citet[][Table 1 and Figure 1]{page12} for the effective transmission curves and the other properties of the XMM-OM filter passbands. The exposure times per filter are summarised in Table \ref{tabobsid}, together with the pointing coordinates and the position angles of the observations.

\begin{table*}
\caption{Detections and sources from individual observations and stacked images, and final contents of the catalogue}             
\label{tabtotdetections}      
\centering                          
\begin{tabular}{l r r r r r r r}        
\hline\hline                 
  & UVW2 & UVM2 & UVW1 & U & B & V & Tot. \\    
\hline                        
   a) Total detections from individual observations & 1125 & 1408 & 7437 & 13468 & 5975 & 3182 & 18519 \\      
   b) QF0 detections from individual observations & 832 & 1164 & 6134 & 9434 & 4225 & 2163 & \\
   c) Sources with QF0 detections from individual observations & 205 & 260 & 1060 & 1708 & 1166 & 683 & \\
   d) Sources with QF0 detections from stacking & 159 & 284 & 1325 & 2041 & 1186 & 415 & \\
   e) Subset of d), without counterparts in the omcat-catalogue & 80 & 138 & 561 & 863 & 645 & 205 & \\
   f) Total sources before validation of the catalogue & 285 & 398 & 1621 & 2571 & 1811 & 888 & 4478 \\
   g) UV sources before validation of the catalogue & 285 & 398 & 1621 & 1075 & 692 & 433 & 1844 \\
   h) UV sources in the XMMOMCDFS catalogue & 157 & 301 & 1121 & 954 & 632 & 403 & 1129 \\
   i) XMMOMCDFS sources from stacking & 46 & 113 & 348 & 66 & 119 & 78 & \\
\hline                                   
\end{tabular}
\tablefoot{a) and b) respectively refer to the total detections and the detections without bad quality flags (QF0) obtained by the standard data processing of the XMM-OM observations that compose the XMM-CDFS Deep Survey. c) the number of individual objects for each filter with at least one QF0 detection from individual observations (omcat-catalogue sources). d) and e) individual objects detected without bad quality flags from stacked images and subsets of these sources that are not included in the omcat-catalogue. f) total number of sources for each filter before the validation of the catalogue (Section \ref{vali}) and before selecting the objects with UV measurements; f) equals c) + e). g) is the subset of f), containing the sources with at least one UV measurement. h) refers to the final contents of the XMMOMCDFS catalogue, after the validation process. i) indicates the number of XMMOMCDFS sources for each filter contributed by stacking and without counterparts from individual observations. The last column is empty for b), c), d), e), and i), because the selections of QF0 detections and of sources from stacking without individual observation counterparts have been performed separately for each filter.}
\end{table*}

The archival XMM-OM data, available from the XMM-Newton Science Archive\footnote{http://xmm.esac.esa.int/xsa/} (XSA) and from the NASA High Energy Astrophysics Science Archive Research Center\footnote{http://heasarc.gsfc.nasa.gov/db-perl/W3Browse/w3browse.pl} (Heasarc), have been re-processed through the standard SAS fixed runs of interlinked tasks described by \citet{page12}\footnote{Unlike the XSA and Heasarc pipeline data, however, the photometric uncertainties are computed by applying a binomial distribution, because, as demonstrated by \citet{kuin08}, these are the correct statistics for the detectors of XMM-OM \citep[microchannel plate intensified CCDs,][]{ford00}}, with constant values of significance and sigma\footnote{The significance is defined as the ratio between the source counts and the respective root mean square background fluctuations, while the parameter sigma controls the minimum detection threshold (all the pixels with values greater than or equal to background value + sigma $\times$ background noise are considered part of a possible source). The standard values are minimum significance=3 and sigma=2.}. The total source list of an XMM-OM observation is composed of i) the detections from its individual exposures and ii) the detections from the sky-image built by mosaicking and stacking the exposure frames for each of the filters, without detections from individual exposures. The new detections from the stacked sky-image are not corrected by SAS for coincidence loss; however, since they are very faint, the correction would be negligible. During the SAS data processing, possible spurious detections can be caused by scattered-light artefacts (smoke rings, diffraction spikes, etc.), and the photometry estimates of true detections may be affected by contaminations, such as the presence of bad pixels. All the dubious detections are automatically marked by specific bad quality flags \citep[QF,][]{page12}, which are summarised for each object (and for each filter) by an integer number; QF=0 for detections whose photometry is to be considered accurate (thereafter we use the acronym ``QF0'' for these detections).

 Matching the total source lists of each XMM-CDFS Deep Survey observation through the SAS package {\it omcat}, which identifies multiple detections of individual objects within 2 arcsec (or three times the position uncertainty), a data set of 18519 measurements is achieved for a total of 4365 sources. It includes all the standard detections obtained for each of the XMM-OM observations that compose the survey. The total and QF0 detections for each XMM-OM band are reported in Table \ref{tabtotdetections} (rows ``a)'', ``b)'', and ``c)''); the incidence of bad quality flags (QF$>$0) ranges between 17\% and 31\%.

%

\subsection{Data processing and stacking}\label{stac}

\begin{table*}[t]
\caption{Cross-identification flags (CIF)}             
\label{tabflag}      
\centering                          
\begin{tabular}{c p{16cm} r}        
\hline\hline                 
CIF & Definition of the flag & N \\    
\hline                        
   0 & Confirmed XMM-OM detection with unique EIS/COMBO-17 cross-identification & 988 \\      
   1 & Reliable XMM-OM counterpart of an EIS/COMBO-17 source with multiple XMM-OM cross-identifications & 35 \\
   2 & Confirmed UVW2/UVM2 detection without UVW1 and U counterparts & 8 \\
   3 & Ambiguous XMM-OM counterpart of an EIS/COMBO-17 source with multiple XMM-OM cross-identifications & 38 \\
   4 & Ambiguous EIS/COMBO-17 counterpart of an XMM-OM source with multiple EIS/COMBO-17 cross-identifications & 4 \\
   5 & Dubious UVW1 detection without U counterpart or potentially wrong cross-identification in the EIS and/or COMBO-17 survey & 56 \\
\hline                                   
\end{tabular}
\end{table*}

Starting from the original catalogue of detections from individual observations, we discarded the QF>0 data for each filter to eliminate potentially spurious or inaccurate photometric measurements. Then, we computed the standard, unweighted average astrometry and photometry of individual objects, matching the possible multiple detections within a maximum separation of 1.5 arcsec, also considering the position uncertainties ({\it Sky with errors} match criterion), by means of the Virtual Observatory application TOPCAT\footnote{http://www.star.bris.ac.uk/$\sim$mbt/topcat}. This cross-correlation criterion (with 1.5 arcsec maximum separation) is very similar to the internal match algorithm developed within the SAS packages (with a 2~arcsec maximum separation), and used before for identifying multiple detections of each source in joining the individual observation source lists with {\it omcat} (sec. \ref{sas}). Thereafter, these catalogues of QF0 detections and average measurements for each filter are indicated as a whole as the ``omcat-catalogue''.

To let fainter sources emerge from the background above the significance threshold, we carried out the stacking of the exposures, per filter, for each of the groups of consecutive observations, by running the {\it ommosaic} package. Subsequently, we ran the sequence of {\it omdetect}, {\it ommag}, {\it omqualitymap}, and {\it omsrclistcomb} tasks for performing the source detection (again adopting the standard sigma and significance thresholds) and the astrometry correction, which is automatically carried out by SAS by comparing the XMM-OM coordinates with the USNO-B1.0 source catalogue \citep{mone03}. We did not stack all the available  exposures as a whole, because the random background fluctuations can be magnified and overtake the cumulative source counts. By adding the image frames of each group separately, instead, it is possible to detect faint sources with accurate photometry: the position angle does not change, and the scattered-light artefacts are confined to limited regions of the field of view.  Furthermore, cataloguing all the sources directly from the stacked images is not feasible: no coincidence loss corrections can be performed on stacked frames, and the photometry of brighter objects cannot be considered reliable in the subsequent detection process.

For integrating the omcat-catalogue with the deeper detections from the stacked images, we extracted the QF0 measurements separately for each filter, and we computed the average astrometry and photometry (row ``d)'' of Table \ref{tabtotdetections}). Then, for each of the filters, we matched the stacked image source lists with the corresponding omcat list of QF0 detections. We carried out the match again through the TOPCAT {\it Sky with errors} criterion, with a match radius of 1.5 arcsec, and in this way we selected the new detections from the stacked images without individual observation counterparts (row ``e)'' of Table \ref{tabtotdetections}). We averaged their coordinates and photometric estimates. (A few objects are detected twice within two or more stacked images of the same filter.) The source-lists that resulted from the concatenation of the omcat sets of individual sources and the new detections from stacking represent the corresponding deep catalogues of XMM-OM in the CDF-S (row ``f)'' in Table \ref{tabtotdetections}). Matching them by coordinates, and finally selecting the objects detected in at least one of the UV bands (either in individual observations and/or stacked images), we obtained a source list of 1844 UV objects (row ``g'' of Table \ref{tabtotdetections}). It is important to note that some of the new detections from stacking in one or more of the filters correspond to objects already included in the omcat-catalogue, because detected in other bands. Therefore, the stacking partly provides detections of new sources without counterparts from individual observations, and partly contributes additional photometric measurements of omcat-catalogue objects. After validating the catalogue (see Section \ref{vali}), with reference to the three UV filters, the contribution of the stacking will amount to $\sim$31\% of new sources without counterparts from individual observations, while providing additional photometry for another $\sim$12\% of sources.

%

\subsection{Validation of the catalogue}\label{vali}

\begin{figure}
   \centering
   \includegraphics[width=\hsize]{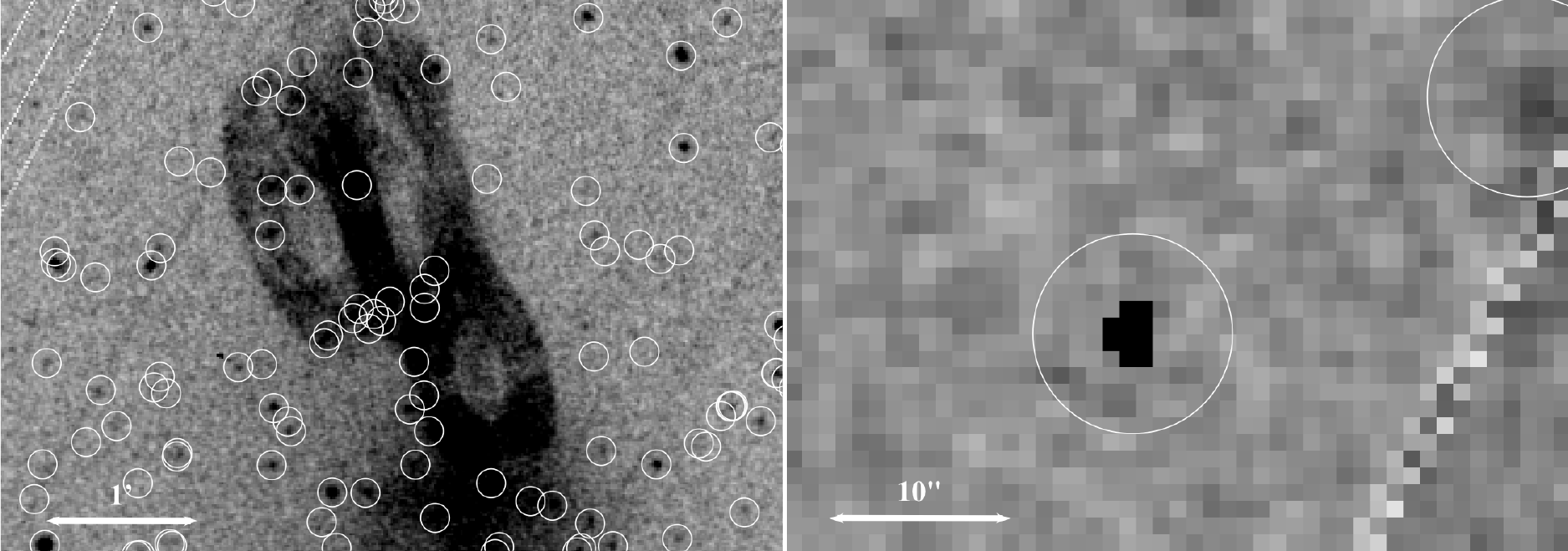}
   \includegraphics[width=\hsize]{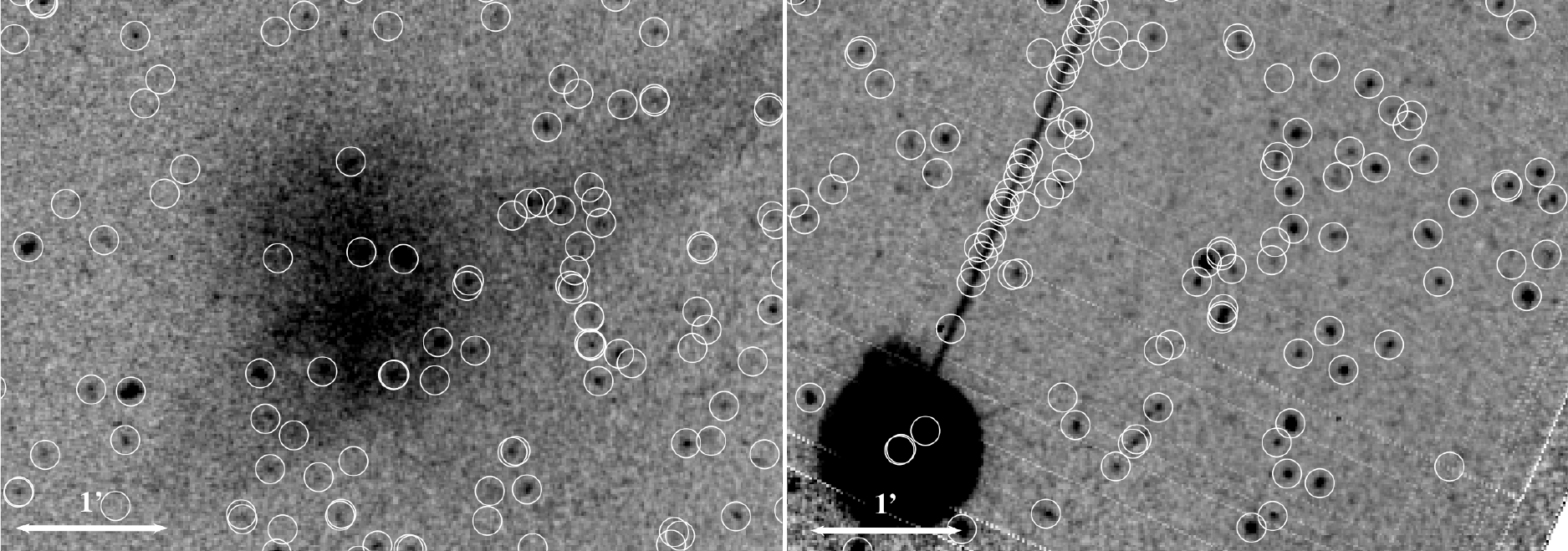}
\caption{Examples of artefacts into the XMM-OM images. Top left panel: internal reflections within the telescope; top right panel: cosmic ray; bottom left panel: central enhanced region; bottom right panel: readout streak. White circles refer to the detections before validating the catalogue. All the images are extracted from UVW1 exposures. Bright stars have not been masked.}
\label{figartefacts}
\end{figure}

Despite the selection of QF0 measurements --- both from the source lists of individual observations and from stacking --- the catalogue still contains potentially false detections. This contamination is pointed out by the subsequent clues: i) a considerable number of sources (367 out of 1844, about 20\%) are detected in only one observation and in a single UV band, even though their magnitudes appear to be relatively bright, revealing the possibility of coincident events among the detections (e.g. cosmic rays, which show a typical square structure of 2$\times$2 very bright pixels); ii) the spatial distribution of the detections partly reflects the distribution of the main artefacts in the images, such as the readout streaks due to a bright star within the area covered by the XMM-OM observations (see Figure \ref{figartefacts}). In other words, while some of the detections marked by bad QFs (QF>0) could be considered reliable and their photometric measurements only marginally inaccurate \citep{page12}, the SAS algorithms of validation do not prevent completely spurious detections to be recognised as true, so included in the catalogue. This occurs especially if, within individual exposure frames, the artificial background fluctuations related to extended artefacts (smoke rings, readout streaks, the central enhanced region of the images) appear to be rarefied, and incidentally present well-defined source-like brighter pixel clusters.

To validate the detections, the cross-identification with sources catalogued by given reference surveys of the CDF-S becomes necessary. For this purpose, we chose the ESO imaging survey \citep{arno01} and the COMBO-17 survey \citep{wolf04, wolf08}, because they are supposed to be complete at the limiting magnitudes that can be reached by XMM-OM in the optical (and UV) bands. In fact, before validating the catalogue, the detection limits in the most sensitive XMM-OM bands (UVW1 and U) are $m$$=$24.7 (see also Figure \ref{figmag} for the UV magnitude distributions after the validation of the catalogue). Moreover, the EIS and COMBO-17 surveys are characterised by a very good astrometric precision better than 0.10 and 0.15 arcsec, respectively. The COMBO-17 catalogue also provides useful photometric classifications (and redshift estimates) by fitting the low resolution spectral energy distributions (SEDs) with specific templates for AGN, galaxies, and stars. We note that this validation process does not account for possible sources whose emission peaks in the far-UV domain, without counterparts in the EIS/COMBO-17 surveys. We estimate that, validating our catalogue with the GALEX observations in the CDF-S, some tens up to one hundred sources without optical cross-identifications (depending on detailed validation after visual inspection on the images) would be added. We chose to only refer to the EIS/COMBO-17 surveys for validating the XMM-OM detections, to take advantage of the classifications provided by the reference optical surveys. We also notice that the validation procedure can lose some possible transient sources not present at the epochs of the reference ground-based surveys.

The cross-correlation procedure adopted for the validation of the catalogue --- and for identifying the sources within other CDF-S surveys, see Section \ref{stat} --- is very standard. After matching the catalogue with the given reference source-list by coordinates, we repeat the cross-correlation with a set of false coordinates, which were obtained by adding an offset of 30 arcsec to the coordinates of the reference survey. Then, by comparing the histograms of the separations, we choose the best match radius for minimising the rate of false matches without losing too many true counterparts: this is determined to be 1 arcsec for the EIS and COMBO-17 surveys. The probability of false matches is estimated through the ratio between false and true cross-identifications within the match radius; the possible offset in the distribution of separations of the true counterparts, instead, indicates a possible bias in the astrometry of the catalogues. The possible incompleteness is estimated as the ratio between the residual matches within the tail of the distribution, beyond the chosen radius, and the total number including such matches.

After selecting the best match radius, the cross-correlation is repeated in the {\it all matches} mode, for determining possible double or multiple counterparts. The choice of the most likely cross-identifications has been generally performed by selecting the counterparts at the lower distance. However, matching the XMM-OM source-list with the EIS and COMBO-17 surveys, visual inspections of the images and photometric considerations have been also carried out. We define a system of cross-identification flags (CIF), reported in Table \ref{tabflag}, to account for the reliability of the identifications.

By applying the procedure described above, the sources farther than 1 arcsec from the EIS or COMBO-17 counterparts are discarded, and a total of 1210 EIS and 1251 COMBO-17 counterparts are identified in the {\it all matches} mode. The probabilities of false matches correspond to 4.1\% and 5.5\%, respectively, while the maximum incompleteness is estimated at 6.1\% and 2.7\%, respectively. The cases of double or multiple XMM-OM cross-identifications of given EIS/COMBO-17 objects are 73; among them, 35 counterparts are clearly recognisable (CIF=1), because the other detections are spurious events (cosmic rays), while 38 are the ambiguous identifications. Only one of the counterparts is included in the main catalogue for each of these sources --- the most likely in terms of angular distance --- and is marked with CIF=3. The discarded counterparts are, however, listed in the supplementary catalogue, described in Section \ref{columns}, together with the main catalogue. Four XMM-OM sources, instead, have two EIS/COMBO-17 counterparts: again, the most reliable of them is selected, but these identifications have to be considered ambiguous --- the astrometric precision of the USNO-B1.0 source catalogue, used in SAS for the astrometry correction, is of about 0.2 arcsec, and the systematics in the distortion map currently turn out to be about 0.7 arcsec rms \citep{tala11} --- and they are marked by CIF=4. A note on the discarded identification is reported in the main catalogue (see Section \ref{columns}).

Visual checks of the images, intended to confirm potentially dubious detections, were also carried out. Spurious detections can be hidden in objects detected in the UVW2 and/or UVM2 bands without measurements in the UVW1 and U filters, and UVW1 sources without U counterparts, because the U sensitivity is higher with respect to the UV filters, and the UVW1 filter is more sensitive than the higher frequency bands. However, the U and UVW1 detections can be missing because of the selection by filter of QF0 measurements: 8 out of 17 UVW2/UVM2 detections without UVW1/U counterparts are real (CIF=2); similarly, 169 out of 239 UVW1 detections without U measurements are real. UVW1 sources that turned out to be clearly spurious (false events, residual detections on scattered-light artefacts) were discarded from the catalogue, while some dubious cases were marked with CIF=5.

 We also checked the cross-identifications that present suspiciously faint photometric measurements (R>24, U'>25; the R magnitudes are provided by both the EIS and the COMBO-17 survey, and U' is the EIS U350/60 WFI filter) or uncommon colours. Very faint optical magnitudes from the EIS/COMBO-17 counterparts, or strong differences in the XMM-OM and EIS/COMBO-17 photometric measurements between nearly coincident wavelength bands may indicate potentially wrong cross-identifications, but also possible physical reasons can account for such measurements (e.g. variability); consequently, the ambiguous identifications are visually checked and flagged with CIF=5 if deemed suspect, without being discarded.

The final catalogue is composed of 1129 UV sources. Among them, 1031 objects constitute the subset of verified detections, with reliable EIS/COMBO-17 cross-identifications (CIF$\le$2, thereafter ``good subsample''), while 98 objects (CIF$\ge$3) have to be considered dubious at least in terms of their identifications in the reference optical surveys. In the next section, magnitude distributions, colour-colour diagrams, and photometry comparisons generally refer to the good subsample. We note that the present catalogue is not to be considered complete, owing to the complex selection that includes photometric and cross-identification flags, which produce an incompleteness mainly for the brighter sources, which are more affected by photometric artefacts. Towards the fainter magnitudes, we estimate the approximate magnitude completeness limits as UVW2$\simeq$21.9, UVM2$\simeq$22.1, and UVW1$\simeq$22.8.

%

\section{Presentation of the catalogue}\label{ana}

\subsection{Statistics and complementary information from cross-correlations}\label{stat}

The XMMOMCDFS catalogue, as resulted from the data processing described in the previous section, contains the average astrometry and photometry of the sources detected in at least one of the three UV filters and with quality flag QF=0, confirmed by cross-correlating the EIS and the COMBO-17 surveys and through partial visual screening of the images. The good subsample is defined by the additional requirement CIF$\leq$2. The photometry is computed as the unweighted average of the measurements from individual observations or from the stacked images. The XMM-OM optical average data and the information from the reference surveys (source identifications, COMBO-17 and X-ray classifications, spectroscopic redshifts) are also included. A total of 1121 out of 1129 sources have been detected in the UVW1 band, while the UVM2 and UVW2 photometry is available for 301 and 157 objects, respectively .

\begin{figure}
   \centering
   \includegraphics[width=\hsize]{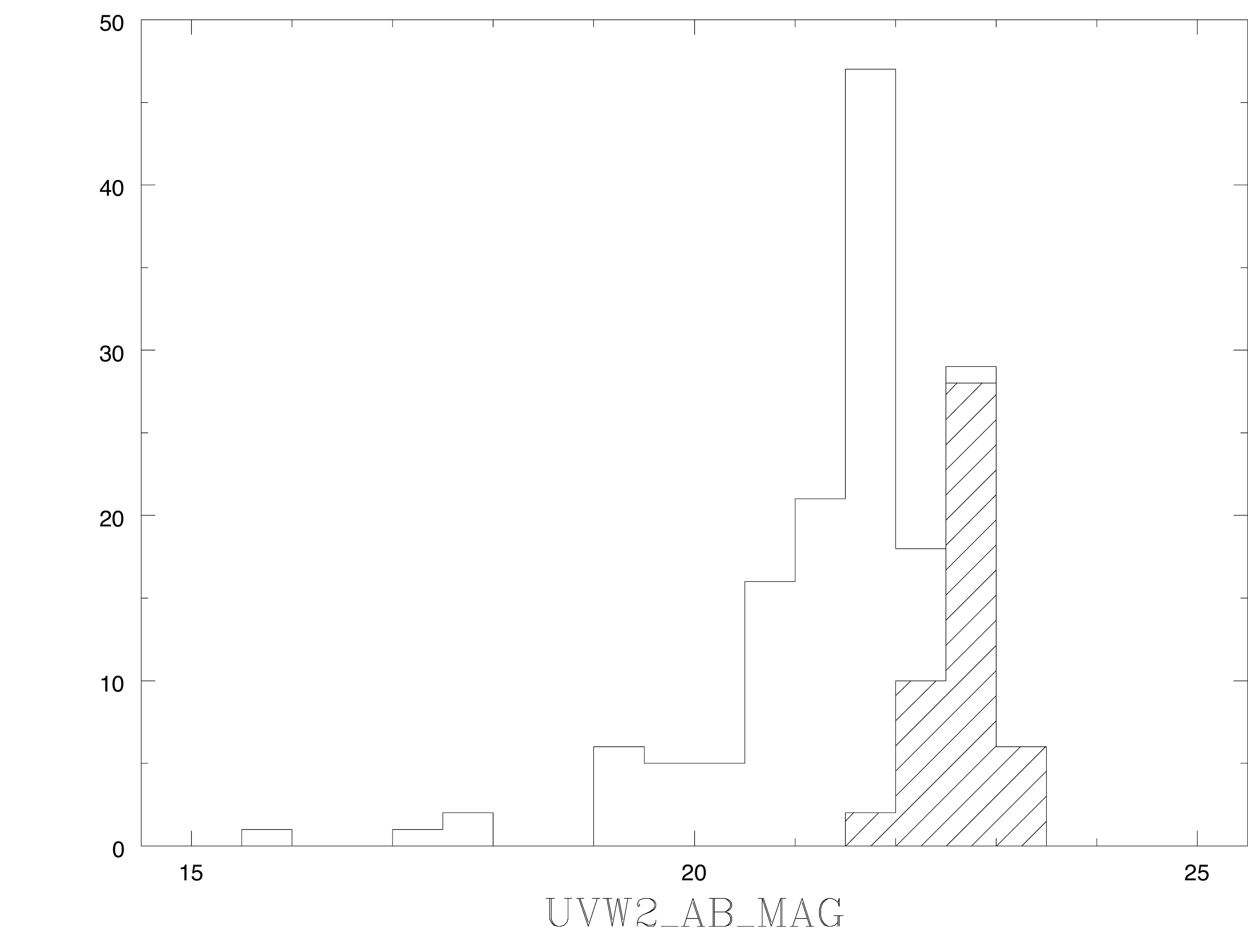}
   \includegraphics[width=\hsize]{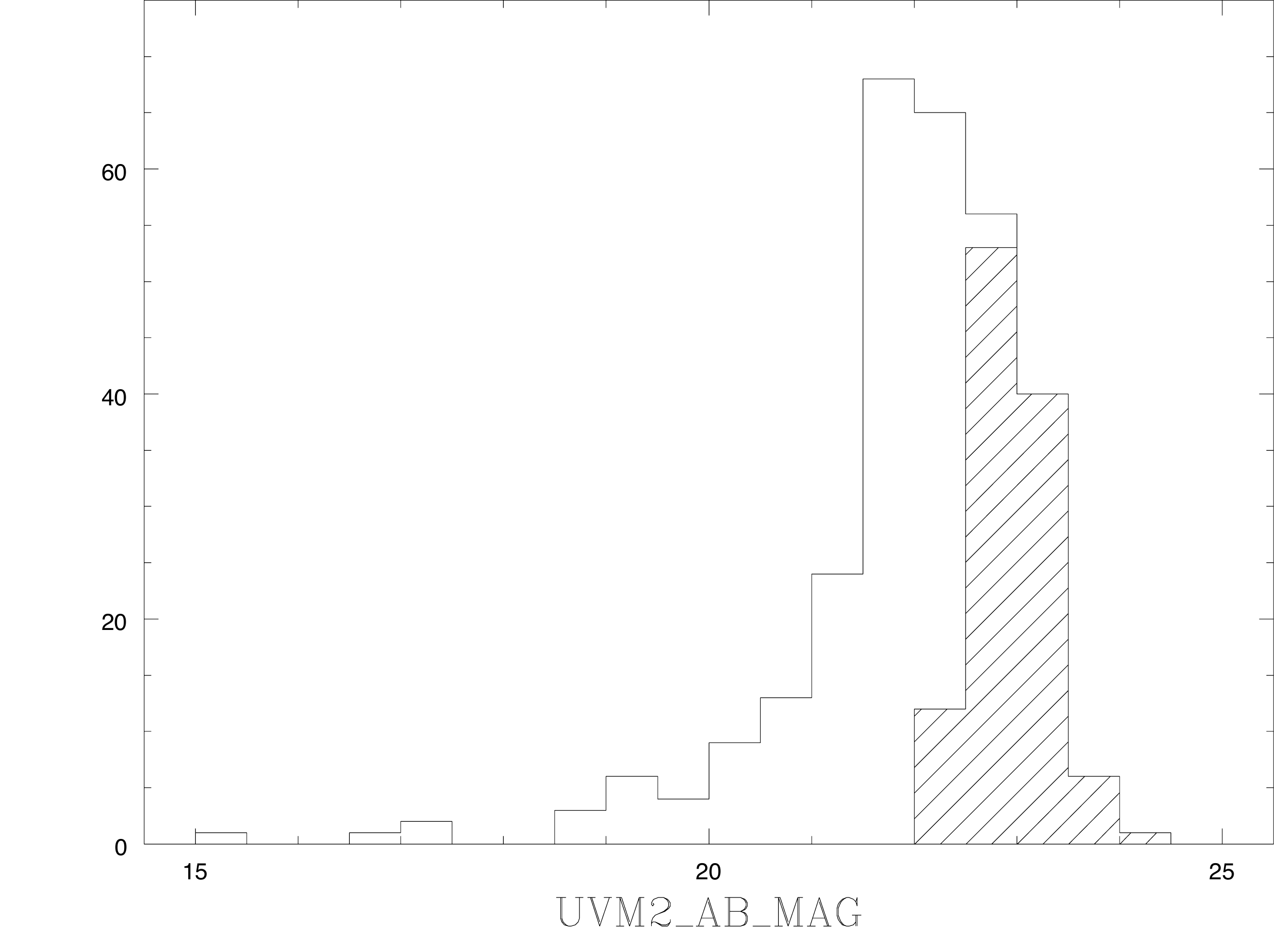}
   \includegraphics[width=\hsize]{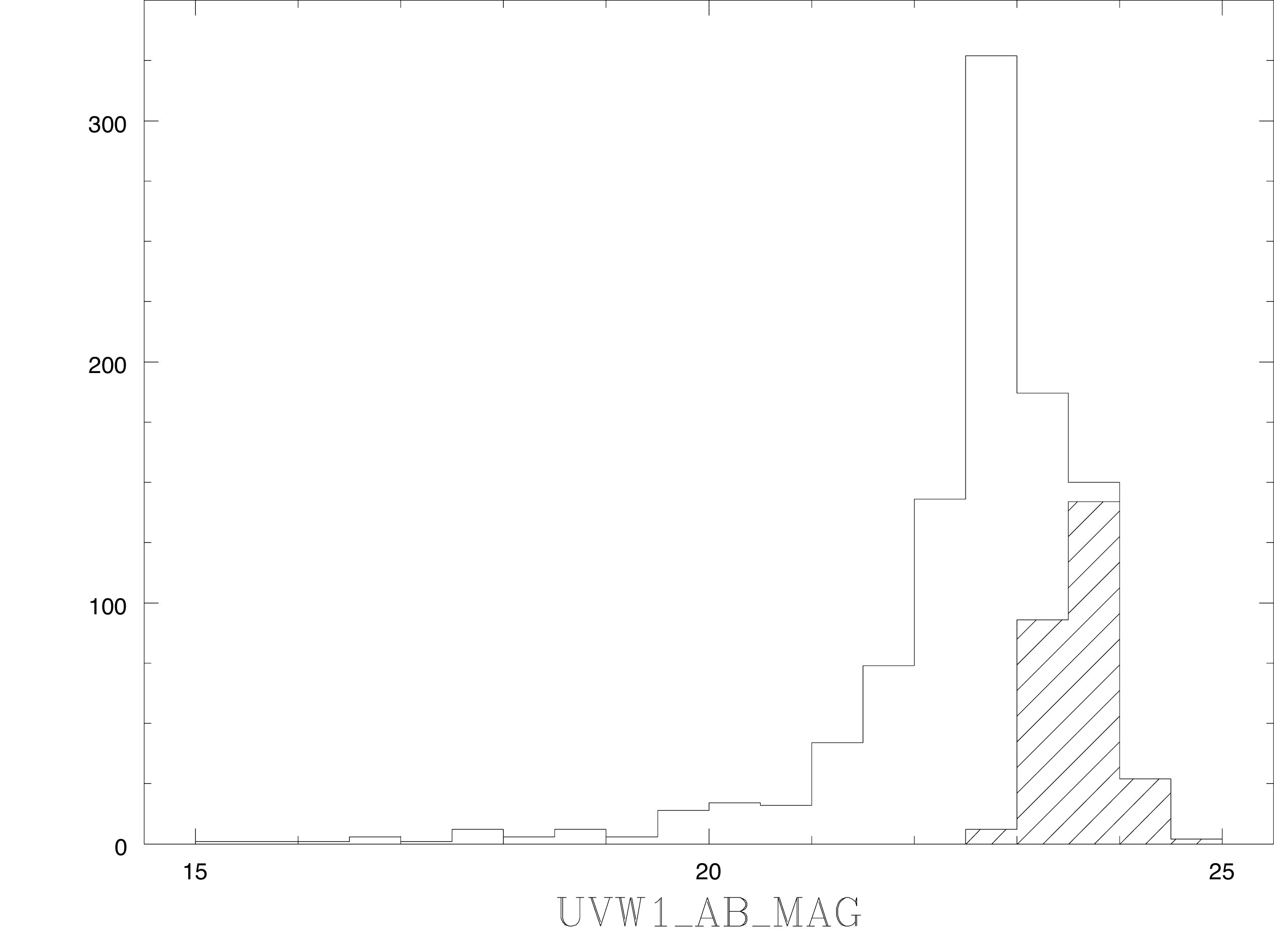}
\caption{Magnitude distributions for the good subsample (empty histograms): upper, middle, and lower panels refer to UVW2, UVM2, and UVW1 bands, respectively. The shaded histograms represent the contribution of the stacking.}
\label{figmag}
\end{figure}

Figure \ref{figmag} shows the magnitude distributions of the good subsample in the UVW2 (upper panel), UVM2 (middle panel), and UVW1 (lower panel) bands, including the contribution by the stacking (shaded parts of the histograms). If compared to the analogous distributions presented by \citet{kunt08} for the OMCat or to the normalised distributions of the XMM-SUSS sources obtained by \citet{page12}, fainter magnitudes are reached both in terms of peaks of the distribution and limiting magnitudes (equal to 23.3, 24.0, and 24.7 for the UVW2, UVM2, and UVW1 bands, corresponding to flux densities of 1.1$\times$10$^{-17}$, 4.9$\times$10$^{-18}$, and 1.6$\times$10$^{-18}$ erg/s/cm$^2$/\AA, respectively), because the exposure times of the XMM-CDFS Deep Survey images are generally longer, and the stacking allows detection of faint sources, increasing the detection limit of each UV band by
$\sim$1 mag. The source detection on stacked images contributes the 29.3\% of the UVW2, the 37.5\% of the UVM2, and the 31.0\% of the UVW1 measurements.

\begin{figure}[t]
   \centering
   \includegraphics[width=\hsize]{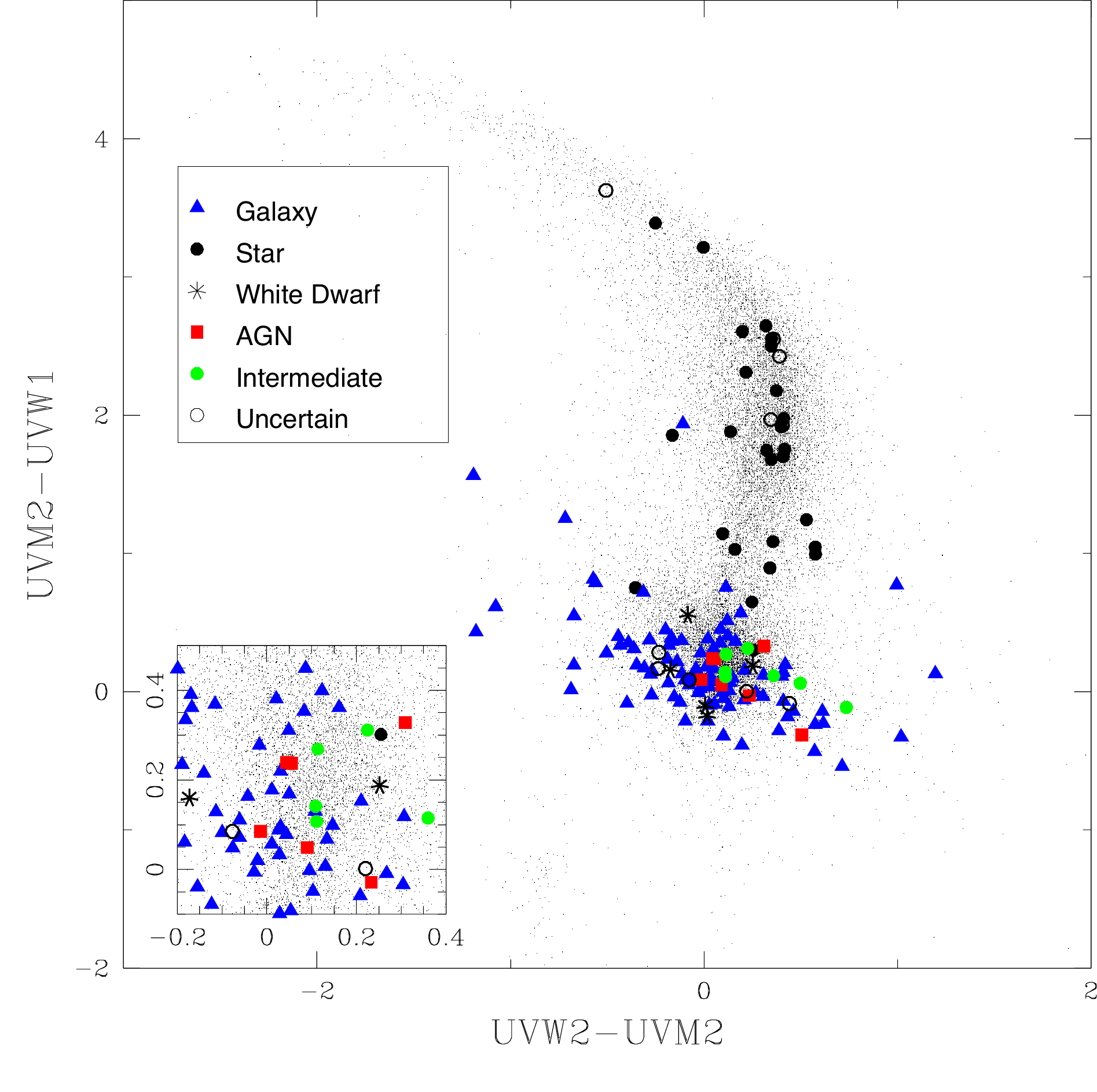}
\caption{UVM2$-$UVW1 versus UVW2$-$UVM2 diagram. The colours of the circles refer to the adopted classification of the objects, based on combining the photometric classifications provided by the COMBO-17 survey and the Chandra 4 Ms source catalogue (see Section \ref{combclas}). Blue triangles indicate the galaxies, stars are represented as black filled circles, white dwarfs as asterisks, red filled squares indicate the AGN, while green filled circles represent the intermediate objects as defined in Section \ref{combclas} (e.g. faint AGN optically diluted by the host galaxy starlight). Black empty circles refer to the sources whose classification is uncertain or unclassified. Symbols are also reported in a legend. The background black dot distribution is the whole XMM-SUSS diagram of \citet{page12}. The most crowded region in the centre of the colour-colour plot is magnified in the small inset.}
\label{figcolourUV}
\end{figure}

\begin{figure}[t]
   \centering
   \includegraphics[width=\hsize]{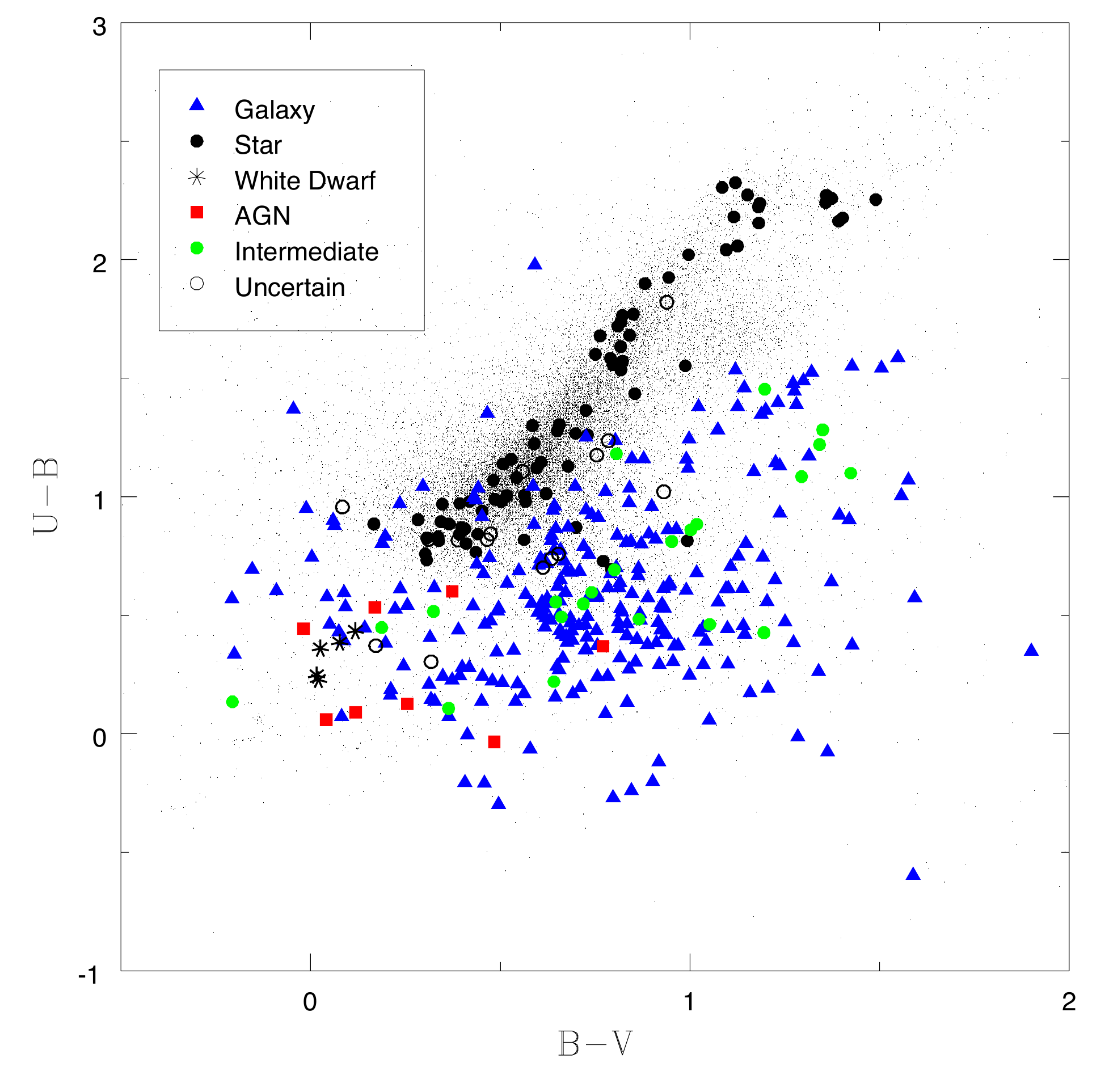}
\caption{U$-$B versus B$-$V diagram. Symbols and colours are defined as in Figure \ref{figcolourUV} and shown in the legend. Small black dots represent the whole XMM-SUSS diagram of \citep{page12}.}
\label{figcolourOptical}
\end{figure}

\begin{figure}[t]
   \centering
   \includegraphics[width=\hsize]{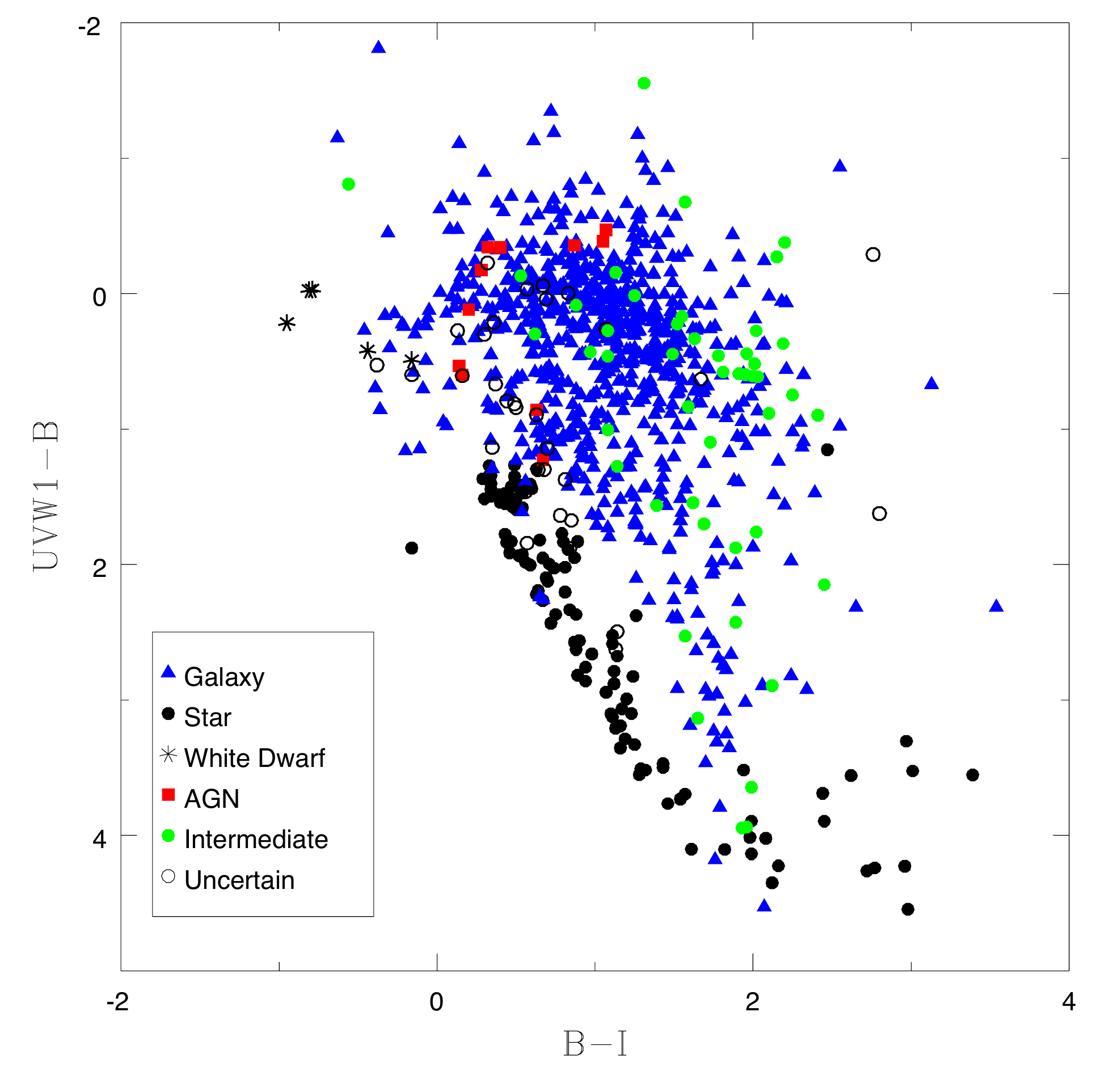}
\caption{UVW1$-$B versus B$-$I diagram, in comparison with similar colour-colour plot by \citet{bian07}, their Figure 5. Symbols and colours are defined as in Figure \ref{figcolourUV} and reported in the legend.}
\label{figcolourEIS}
\end{figure}

Figures \ref{figcolourUV} and \ref{figcolourOptical} represent the UVM2$-$UVW1 versus UVW2$-$UVM2 and the U$-$B versus B$-$V colour plots, respectively. Only the sources detected in all the XMM-OM UV bands (Figure \ref{figcolourUV}) and/or optical bands (Figure \ref{figcolourOptical}) are shown. No corrections for dust extinction or cosmological redshift were applied to produce the colour-colour diagrams. The distributions are presented as functions of the adopted classifications, obtained combining the COMBO-17 photometric classifications and the available estimates of likely source types given by \citet{xue11} --- see Table \ref{tabclassif} and Section \ref{combclas} --- and are compared with the colours of the XMM-SUSS sources \citep{page12}. Figure \ref{figcolourEIS}, instead, is obtained by defining the colours through the UVW1 and the EIS B and I bands, as done for instance by \citet{bian07}, combining GALEX and Sloan Digital Sky Survey (SDSS) photometric measurements in the NUV$-$g versus g$-$i diagrams. In all the colour-colour plots, stellar and galactic loci are clearly distinguishable, even if partly overlapped, and are located in the corresponding colour regions of the comparison distributions and of the models depending on age and temperature \citep[for galaxies and stars, respectively: see][and the references therein]{bian07, bian11, page12}. The subset of ``intermediate'' sources (e.g. optically diluted AGN) is distributed within the entire galactic locus: besides the redshift dependence, this is also due to the optical contamination of the host galaxies (see Section \ref{combclas}).

\begin{figure}[t]
   \centering
   \includegraphics[width=\hsize]{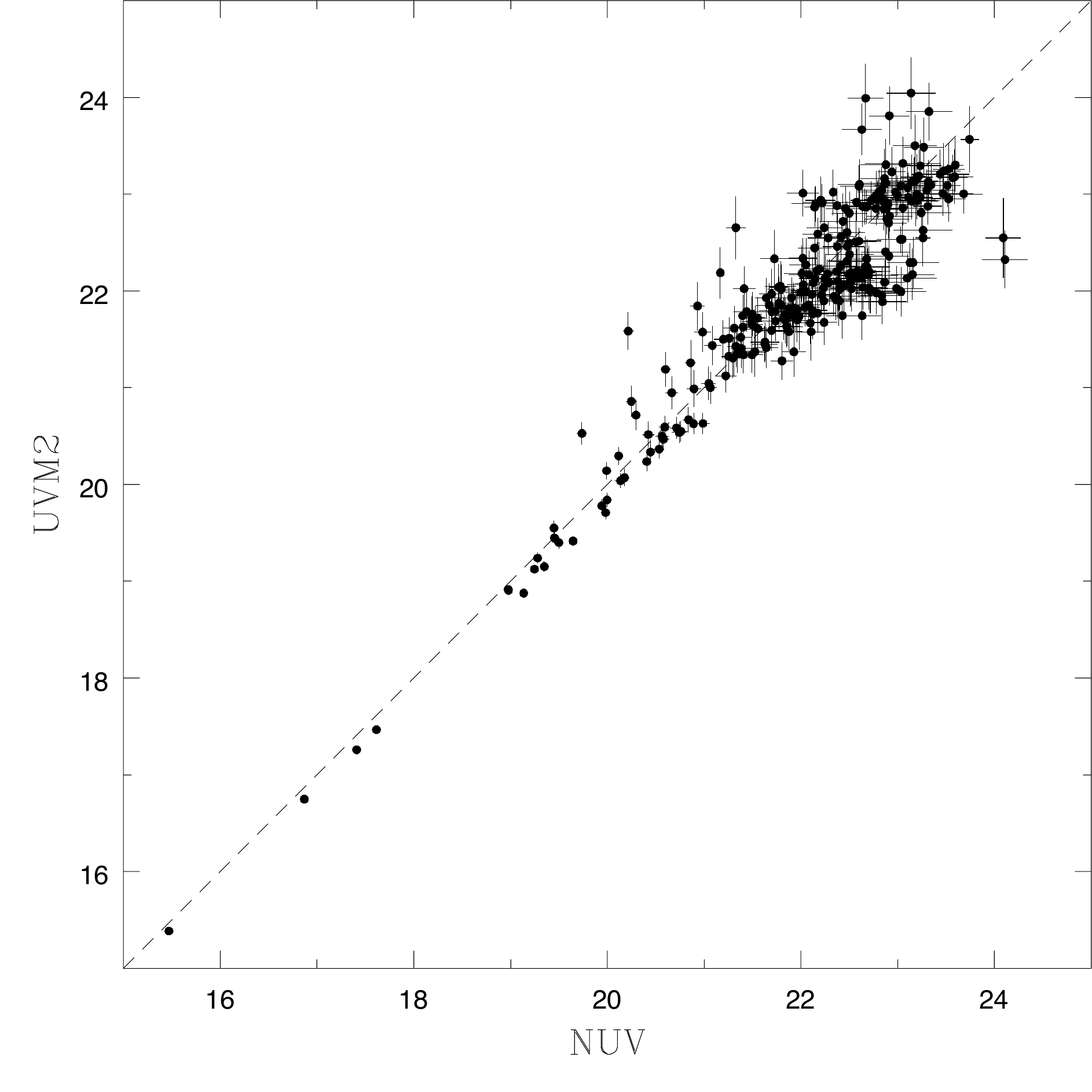}
\caption{XMM-OM UVM2 magnitudes versus GALEX NUV magnitudes diagram. The dashed line indicates the equality of the two magnitudes. Error bars represent one standard-deviation photometric uncertainty.}
\label{figNUVUVM2}
\end{figure}

\begin{figure*}
   \centering
   \resizebox{\hsize}{!}{\includegraphics{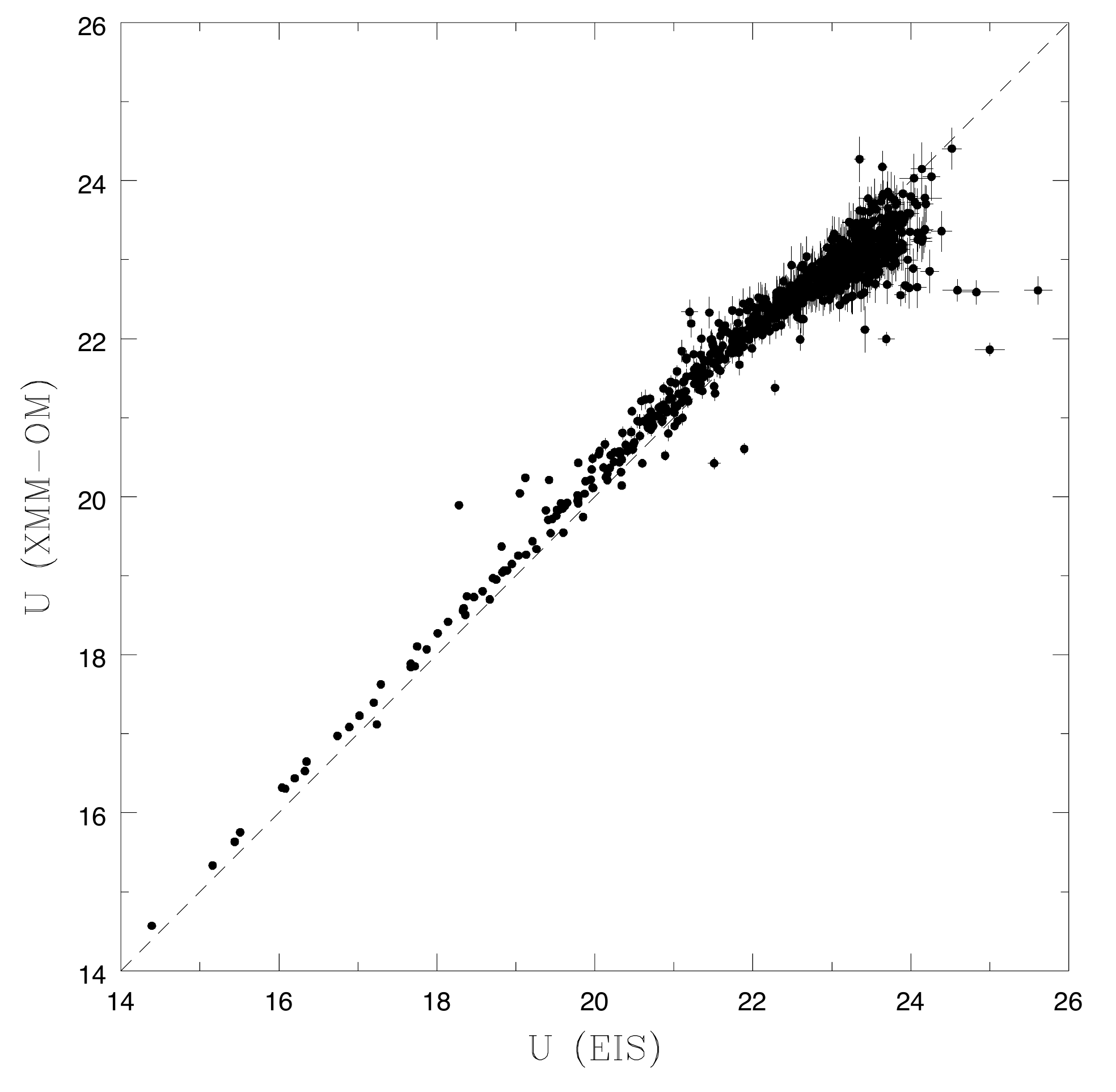} \includegraphics{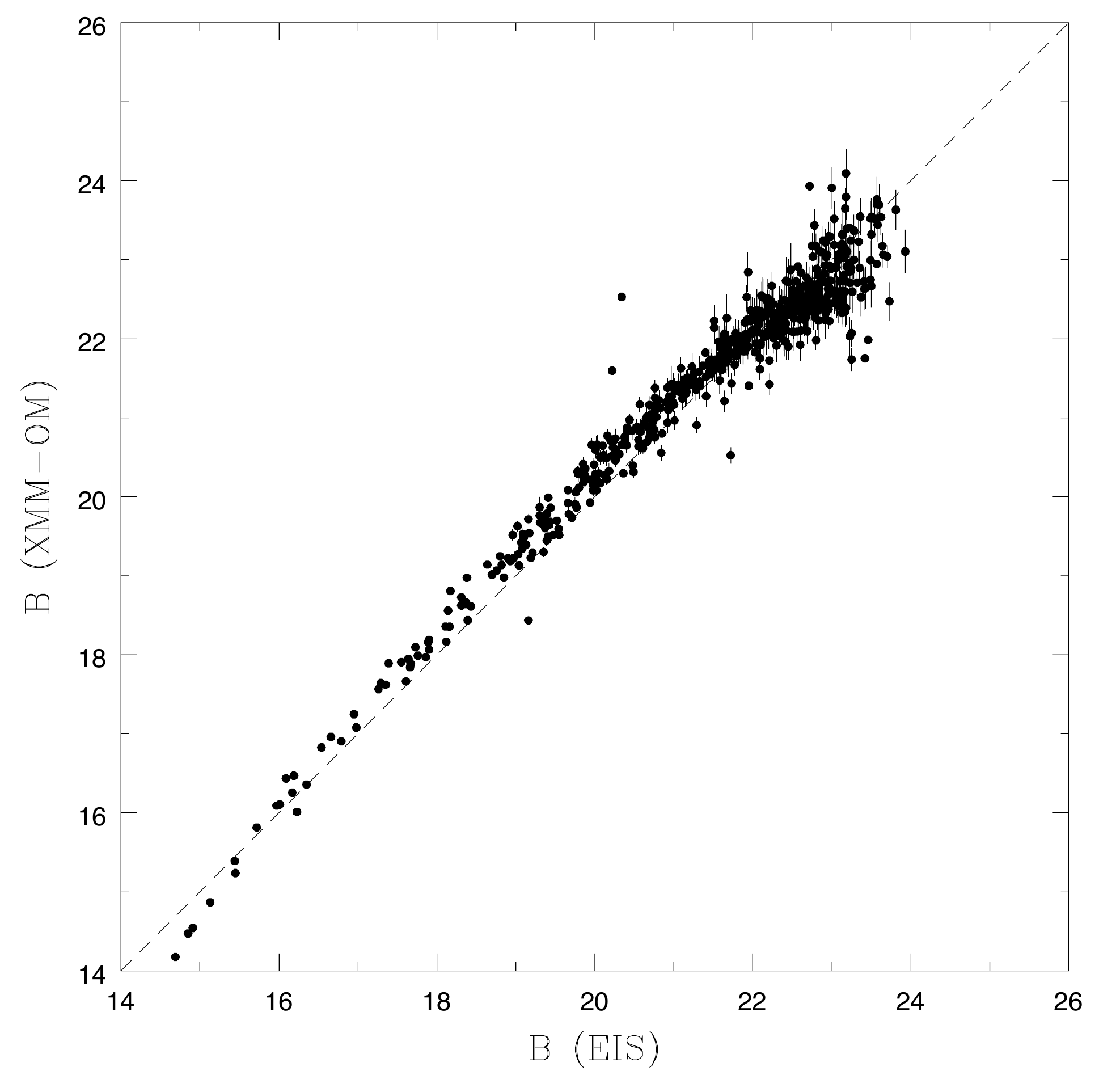} \includegraphics{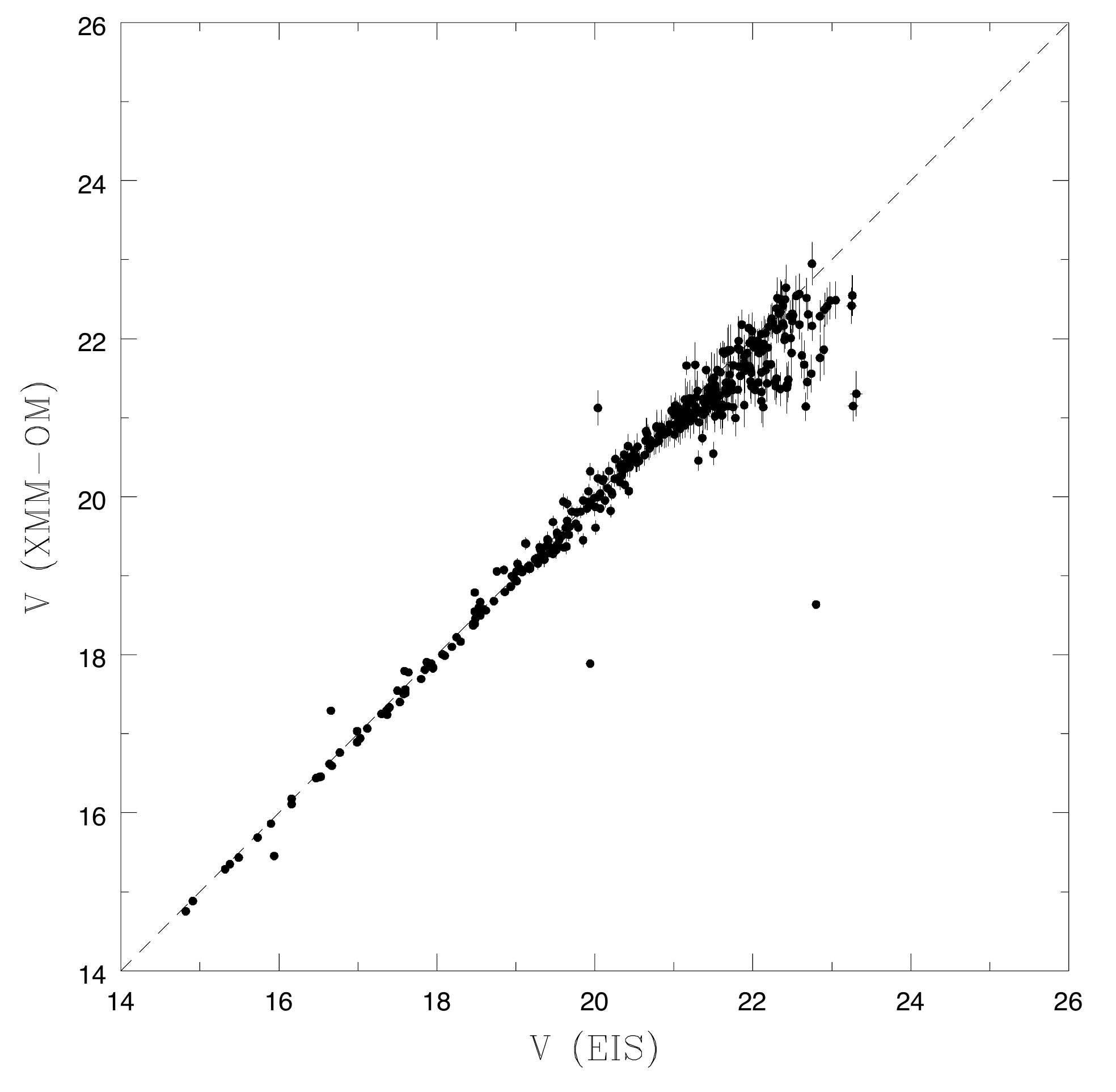}}
\caption{U, B, V comparison of the XMM-OM magnitudes with the corresponding EIS magnitudes. The dashed lines indicate equality of the magnitudes. Error bars represent one standard deviation photometric uncertainties.}
\label{figEISOM}
\end{figure*}

Subsidiary information about the sources were derived by matching the catalogue with other CDF-S surveys, following the cross-correlation procedure of Section \ref{vali}. The GALEX observations provide 957 counterparts within a maximum distance of 2.5 arcsec (877 good sample cross-identifications). The incidence of false matches is estimated to be about 8.9\%, while the incompleteness corresponding to the possible sources farther than the cross-correlation radius is $\leq$5.4\%. There are 13 XMM-OM or GALEX sources with two cross-identifications, and the most likely was attributed for each of them. However, we added a reference to the alternative identification in the main catalogue. We notice that, while it could be expected that all the sources detected by XMM-OM are also found by GALEX, owing to the wider field of view and deeper detection limit, $\sim$15\% of them are missing in the cross-correlation. This could be due to the GALEX photometry in crowded regions, where nearby sources could be interpreted as a merged one with the attribution of wrong coordinates. Examples are reported in \citet{simo14} or \citet{bian14b}. This cross-correlation is useful both for integrating the UV spectral coverage of the sources and for comparing the XMM-OM photometry to the GALEX measurements, in order to examine the calibration of the instrument \citep[see][]{tala11}. Regarding the latter purpose, the most suitable comparison can be made by plotting the XMM-OM UVM2 average photometric measurements as a function of the GALEX NUV photometry (Figure \ref{figNUVUVM2}), because the effective wavelength and the wavelength range of these filters are nearly coincident, even though the two filters have very different shapes. The dispersion in the NUV-UVM2 distribution is partly to be attributed to the relatively high XMM-OM photometry uncertainties at fainter magnitudes, while the possible contamination of false cross-identifications may account for the presence of outliers. As a whole, the UVM2 and NUV photometric estimates are in good agreement, confirming the validity of the XMM-OM calibration.

Similar trends are found by comparing the XMM-OM U, B, and V measurements with the EIS photometry in the corresponding bands (Figure \ref{figEISOM}), which are characterised by almost equal wavelength ranges and effective wavelengths \citep{arno01}. At bright magnitudes, the U and B bands show small systematics, probably because of the differences in the effective wavelengths and shape of the filters.

\begin{table}[b]
\caption{Cross-identifications of the XMM-OM sources in the reference catalogues}             
\label{tabcrossid}      
\centering                          
\begin{tabular}{l r r r r}        
\hline\hline                 
 Survey & Radius & Cross-id & Good & Prob \\    
\hline                        
   GALEX & 2.5 & 957 & 877 & 8.9 \\      
   ACES & 1.0 & 751 & 694 & 1.2 \\
   Master Catalogue & 1.0 & 191 & 173 & 2.5 \\
   Chandra 4 Ms & 2.0 & 164 & 158 & 4.1 \\
   XMM-Newton & 3.5 & 33 & 31 & 12.1 \\
\hline                                   
\end{tabular}
\tablefoot{Radius: match radius of the cross-correlation in arcsec. Cross-ID: total number of cross-identifications. Good: Cross-identifications of good subsample sources. Prob: probability of false matches.}
\end{table}

\begin{figure}
   \centering
   \includegraphics[width=\hsize]{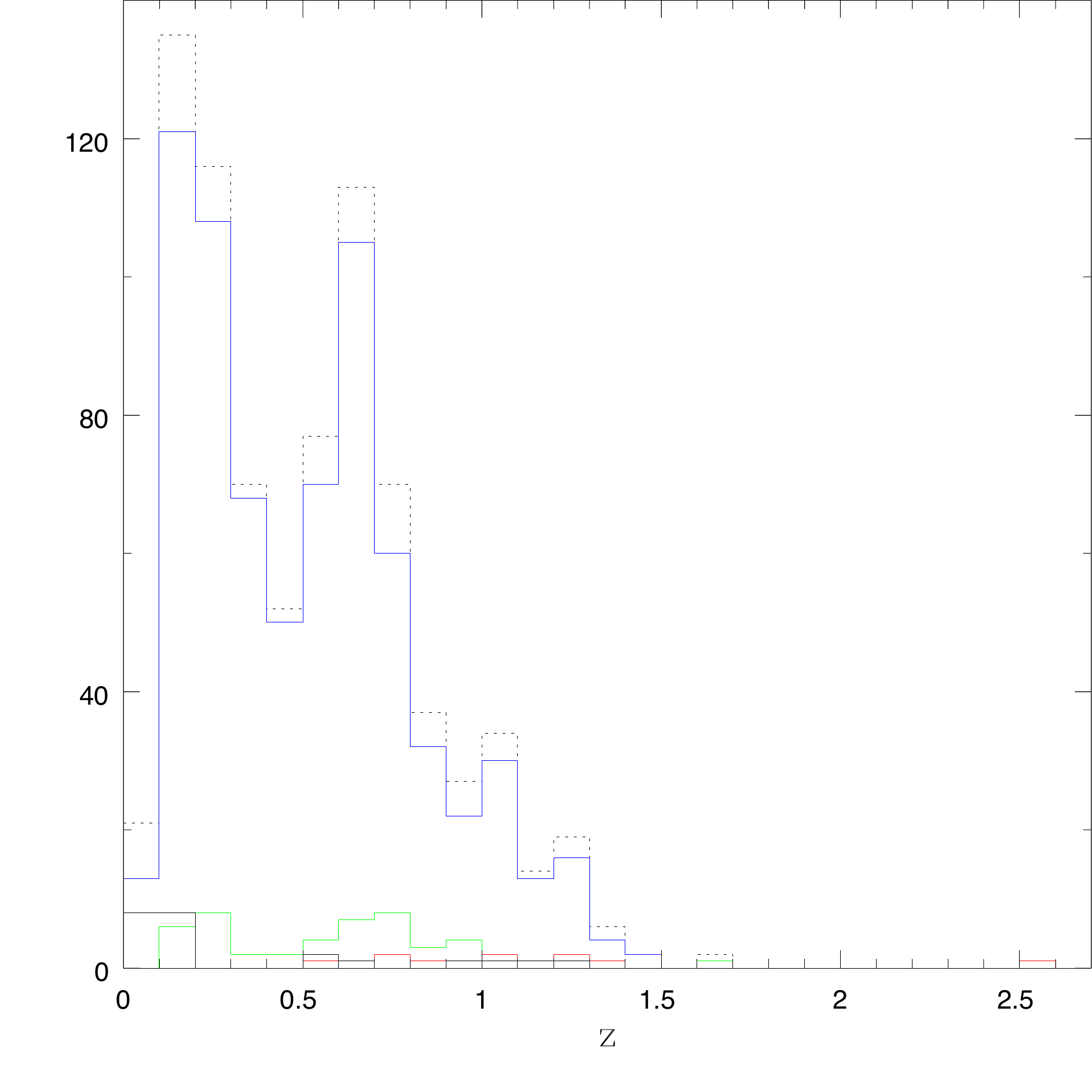}
\caption{Spectroscopic redshift distribution of the extragalactic sources in the good subsample. The dashed histogram refers to the complete subset. Continuous histograms refer to galaxies (blue), AGN (red), intermediate sources (green), and uncertain/unclassified objects (black).}
\label{figredshift}
\end{figure}

\begin{figure}
   \centering
   \includegraphics[width=\hsize]{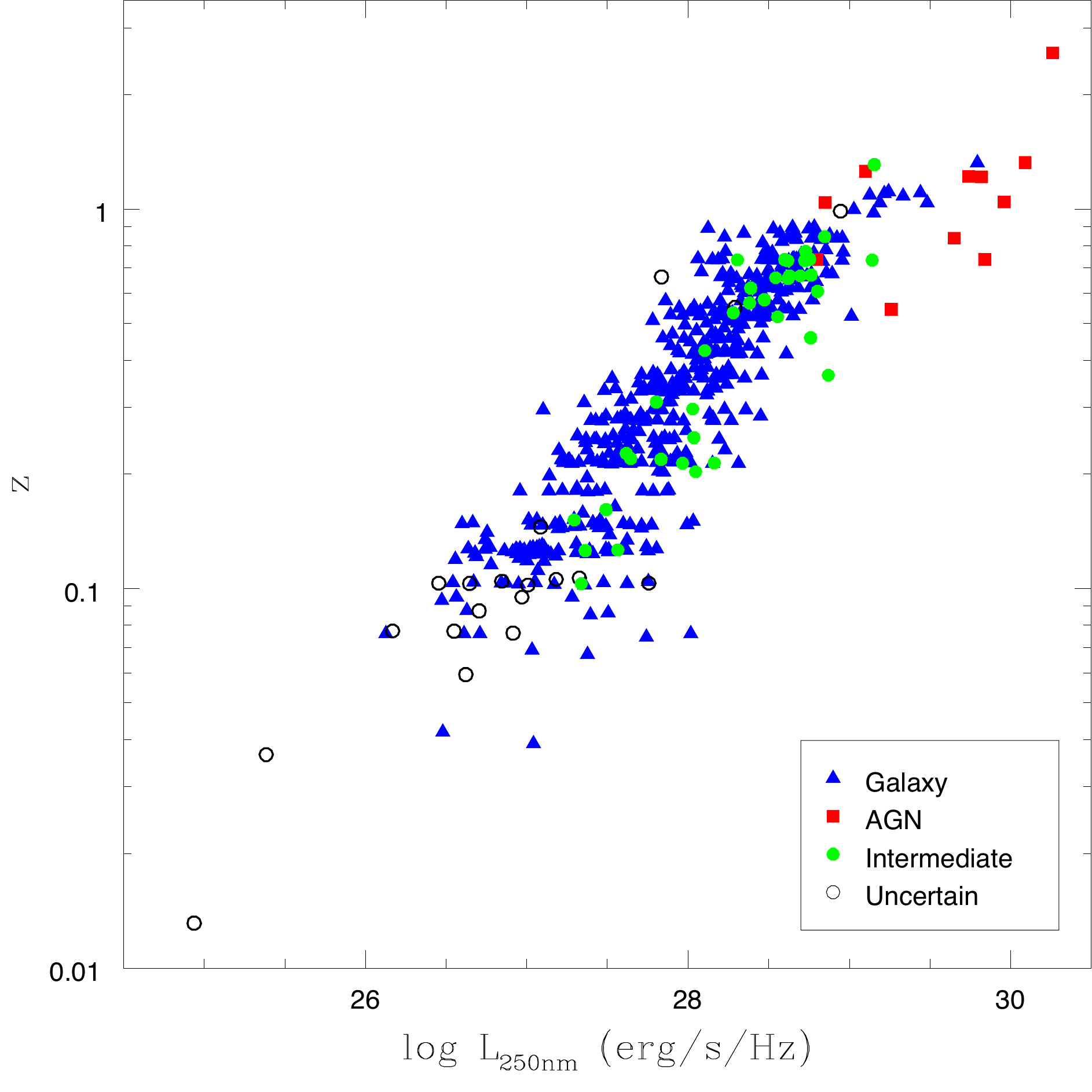}
\caption{ $L_{250nm}$-$z$ plane for the extragalactic objects. Symbols and colours are defined as in Figure \ref{figcolourUV} and reported in the legend.}
\label{figlz}
\end{figure}

While the COMBO-17 survey provides useful photometric estimates of $z$, it is possible to attribute spectroscopic redshifts to the sources of the catalogue by cross-correlating the Arizona CDFS Environment Survey. We obtained 751 ACES counterparts with secure\footnote{We only referred to the estimates denoted by ACES quality code Q equal to -1 (stellar source), 3 (reliable at >90\%), or 4 (reliable at >95\%).} redshift estimate (only one double cross-identification), 694 of which referred to the good subsample, within a maximum distance of 1 arcsec. The possible contamination of spurious matches is 1.2\%, while the maximum incompleteness is estimated to be 4.9\%. The spectroscopic estimates of $z$ for other 191 sources (173 good sample objects) have been extracted from the ESO CDF-S Master Catalogue, version 3.0, considering only the best match for each source. Similar to the ACES counterparts, we considered only the reliable determinations, selecting them by means of the quality flags attributed in each of the references that make up the ESO CDF-S Master Catalogue. The probability of false identifications is 2.5\%. The overall redshift distribution of the extragalactic sources in the good subsample is shown in Figure \ref{figredshift}. The double peak in the redshift distribution is also present in the ACES survey \citep[][their Figure 5]{coop12}, and is partially to be attributed to a large-scale structure in the CDF-S. In fact, \citet{trev07a} and \citet{sali09} found evidence of an over-density at redshift $\sim$0.7, which corresponds to one of our peaks. Figure \ref{figlz} represents the redshift as a function of the specific UV luminosity at 250 nm. In this case, only the sources whose SEDs include at least one measure in the interval 15.00<$\log \nu$<15.16 have been considered, because the uncertainty in extrapolating the specific luminosity at 250 nm ($\log \nu$=15.08) can be considered negligible (see Section \ref{combclas}). Figures \ref{figredshift} and \ref{figlz} refer to extragalactic sources. No unclassified objects have redshifts in the range 0<$z$<0.01, avoiding possible contaminations from stellar radial motions.

 Finally, we cross-correlated the XMM-OM catalogue with the Chandra 4 Ms source catalogue and the 2-10 keV point-source catalogue of XMM-Newton in the CDF-S for identifying the X-ray counterparts of the UV sources. Selecting a match radius of 2 arcsec, we found 164 Chandra cross-identifications, with a 4.1\% probability of false matches, and incompleteness $\leq$4.1\%. The good subsample counterparts are 158. Instead, the cross-correlation with the 2-10 keV point-source catalogue by \citet{rana13} with a match radius of 3.5 arcsec only yields 33 X-ray counterparts (31 for the good subsample) with a 12.1\% probability of false matches and an estimated incompleteness $\leq$13.2\%. The sources detected both by Chandra and XMM-Newton are 27. We notice that, notwithstanding the simultaneity of the X-ray and XMM-OM observations, only 33 out of 339 X-ray sources catalogued by XMM-Newton have UV counterparts in the catalogue. This is expected, because the catalogues by \citet{rana13} refer to the hard X-ray bands (2-10 keV, 5-10 keV), which are well suited to AGN and obscured objects, but less interesting for normal and star-forming galaxies, which constitute the large majority of our catalogue (see Section \ref{combclas} and Table \ref{tabclassif}). Once we take the Chandra 4 Ms catalogue data into account, for instance, 158 sources are detected in the soft-band, 93 of which are galaxies ($\sim$59\%), while only 60 sources are detected in the hard band with only 14 galaxies ($\sim$23\%). Moreover, the small overlap between the XMMOMCDFS and 2-10 keV point-source catalogues is also partly due to the different size of the fields of view (about 140 X-ray sources fall outside the region covered by the XMM-OM observations), and partly has to be attributed to possible biases introduced by the data processing (selecting the QF0 detections and discarding the optical sources without UV measurements, which cause the loss of 35 cross-identifications).

The total statistics of the cross-correlations with the reference CDF-S surveys is summarised in Table \ref{tabcrossid}.

%

\subsection{Combined classification of the sources. X-ray/UV ratio}\label{combclas}

\begin{table}[t]
\caption{Combined classification of the good subsample sources}             
\label{tabclassif}      
\centering                          
\begin{tabular}{l r r r}        
\hline\hline                 
 & COMBO-17 & Chandra & a.c. \\    
\hline                        
   Galaxies & 862 & 91$^a$ & 814 \\      
   AGN & 11 & 61$^b$ & 11 \\
   Intermediate & & & 50 \\
   Stars (and WDs$^c$) & 118 & 6$^d$ & 119 \\
   Uncertain/unclassified & 40 & & 37 \\
    & & & \\
   Total & 1031 & 158 & 1031 \\
\hline                                   
\end{tabular}
\tablefoot{($^a$): two sources with dubious classification --- ``Galaxy (Uncl!)'' --- in the COMBO-17 survey. ($^b$): 50 sources classified as galaxies in the COMBO-17 survey. ($^c$): white dwarfs (WDs) are classified separately only in the COMBO-17 survey. ($^d$): one absent source among the COMBO-17 counterparts. The last column (a.c.) refers to the classification adopted in producing Figures \ref{figcolourUV}, \ref{figcolourOptical}, \ref{figcolourEIS}, \ref{figredshift}, \ref{figlz}, \ref{figalphaox}, \ref{figXR} and contained in column 96 of the catalogue (see Section \ref{columns}). Intermediate sources are the objects classified as AGN in the Chandra 4 Ms source catalogue and as galaxies in the COMBO-17 survey. Objects with dubious classifications --- ``Galaxy (Uncl!)'', ``Galaxy (Star?)'', ``QSO (Gal?)'', ``Strange Object''  --- or absent in the COMBO-17 survey are indicated as uncertain.}
\end{table}

The Chandra 4 Ms source catalogue provides an estimate of the likely source types of the objects, based on revealing possible X-ray emission properties ($L_{0.5-8keV}$$\ge$3$\times$$10^{42}$erg/s; photon index $\le$1.0; $\log (f_X/f_R)$>$-$1\footnote{$f_X$ refers to the flux in one of the 0.5-8 keV, 0.5-2 keV, and 2-8 keV bands}; X-ray emission at least three times higher than the level corresponding to pure star formation\footnote{The X-ray emission attributed to starburst galaxies amounts to 8.9$\times$10$^{17}$$L_R$, where $L_R$ represents the rest-frame 1.4 GHz monochromatic luminosity in units of W/Hz \citep{alex05, xue11}.}) and optical spectroscopic features (e.g. broad emission lines) that characterise the active galactic nuclei \citep{xue11}. Among 158 sources of the good subsample with Chandra X-ray counterparts, 61 are classified as AGN, even if 50 of them are photometrically estimated to be galaxies within the COMBO-17 survey (see Table \ref{tabclassif}). This disagreeing classification likely refers to faint AGN whose optical/UV emission is photometrically diluted by the host galaxy starlight, as suggested by some remarks: e.g. i) most of the sources are optically extended, presenting both EIS and COMBO-17 stellarity indexes $\sim$0 (the COMBO-17 stellarity indexes of 39 objects are 0); ii) the 0.5-8 keV luminosity is <10$^{42}$ erg/s for 23 objects, and in another 7 cases is <3$\times$10$^{42}$ erg/s. Moreover, the available spectroscopic analyses from the literature confirm the double COMBO-17/Chandra classification. Eleven out of 17 cross-identifications in \citet{szok04} are optically classified as ``LEX', which are galaxies that show unresolved emission lines that are consistent with HII region-type spectra, in which the possible presence of the AGN cannot be optically proved, while their X-ray luminosities have to be attributed to the AGN emission (``QSO-1'', ``AGN-1'', ``AGN-2''); analogously. Seven out of 11 counterparts from \citet{trei09} are classified as emission line galaxies (``ELG''), absorption line galaxies (``ALG''), or unknown objects (``UNK''). The dilution of the nuclear emission lines by the host galaxy continuum indeed accounts for the considerable number of relatively faint X-ray sources that are classified as galaxies in the optical follow-up observations \citep{mora02, card07, trei09}.

\begin{figure}
   \centering
   \includegraphics[width=\hsize]{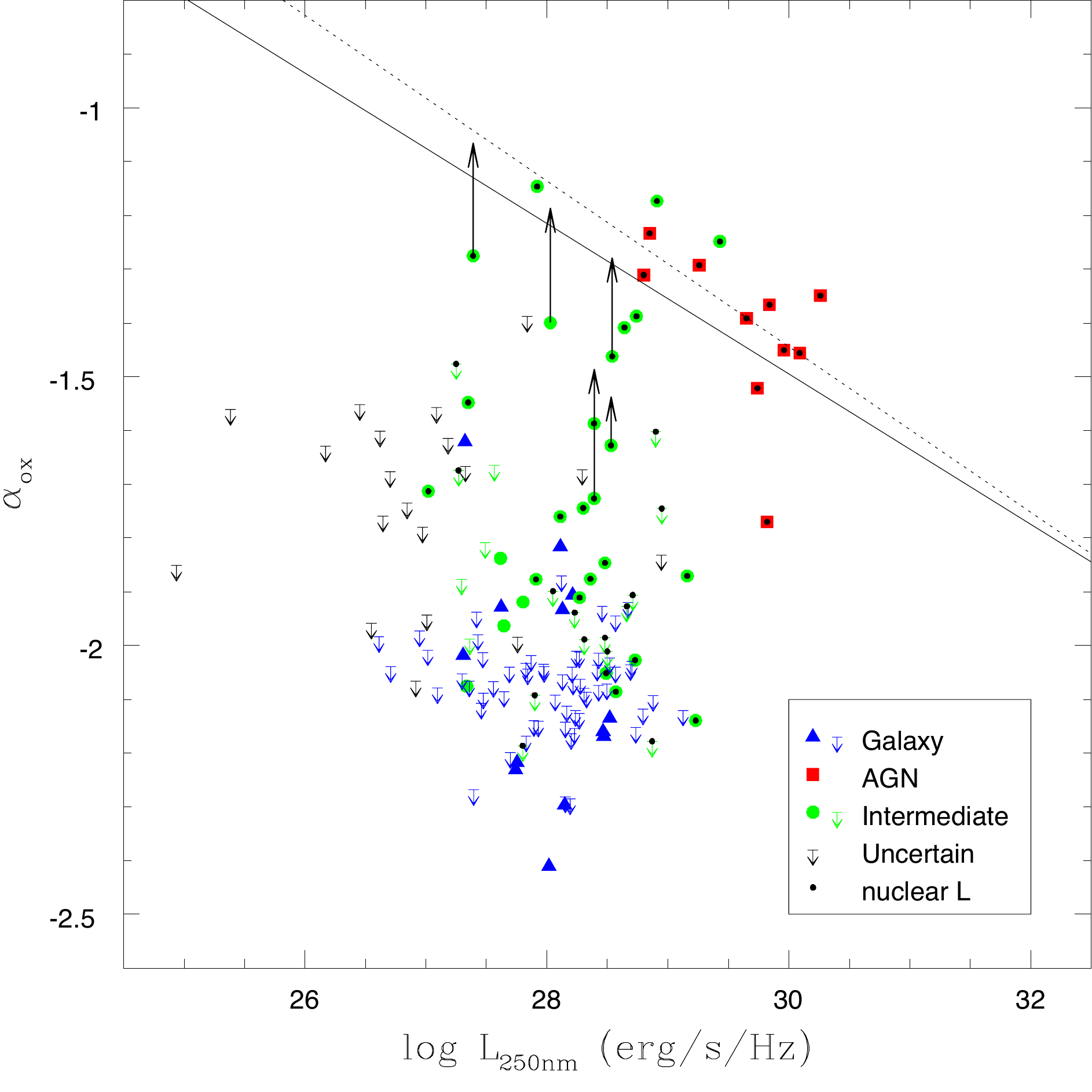}
\caption{$\alpha_{ox}$ versus $L_{250nm}$ diagram. Blue triangles and upper limits represent galaxies, red squares indicate AGN, green circles and upper limits refer to intermediate sources, and black downward arrows represent the upper limits of the Chandra undetected sources. AGN and intermediate sources are marked by small black circles when referring to their nuclear luminosities. All these symbols are also shown in the legend. Green symbols without the black circles refer to the total luminosities of intermediate sources for which it was not possible to determine the nuclear components; a correction performed as in \citet{vagn13} would shift such sources in the upper left direction with slope $\sim$0.38. For some of the intermediate sources, black upward vertical arrows
indicate the correction for the X-ray absorption based on the available X-ray spectra from the XMM-CDFS Deep Survey. The straight lines refer to the best fit $\alpha_{ox}$-$L_{250nm}$ relations by \citet{just07} (continuous) and \citet{luss10} (dotted) for AGN, as reference.}
\label{figalphaox}
\end{figure}

To quantify the X-ray/UV ratio of the objects with X-ray Chandra counterparts, we computed the X-ray-to-optical index, $\alpha_{ox}$$=$$\log(L_{2keV}/L_{250nm})/\log(\nu_{2keV}/\nu_{250nm})$, as a function of the logarithmic UV specific luminosity at 250 nm, as is usually done when analysing samples of AGN \citep[e.g.][]{stra05, just07, gibs08, grup10, luss10, vagn10, vagn13}. To estimate the nuclear contribution of the luminosity at 250 nm, we adopted a two-component fit of the XMM-OM rest-frame SEDs for AGN and part of the intermediate sources. The fitting SED includes a host galaxy starlight component $\propto$$\nu^{-3}$, added to an AGN component shaped as the average SED of \citet{rich06}. The details are described in \citet{vagn13} (see also a similar procedure in \citet{luss10}). For galaxies, intermediate sources whose nuclear emission is found to be totally negligible through the two-component fit, and for uncertain objects, $\log L_{250nm}$ is computed by interpolating or extrapolating the SEDs through a power law of the same slope as the one between the lowest frequency rest-frame luminosity measures. Only the SEDs that include data within the interval 15.00<$\log\nu$<15.16 have been considered in estimating the specific luminosity at 250 nm, to reduce the uncertainty related to the extrapolation method.

The X-ray specific luminosities at 2 keV were computed from the Chandra 2-8 keV fluxes, by also applying a standard power-law k-correction (the values of the photon indexes are provided by the Chandra 4 Ms source catalogue). If the 2-8 keV fluxes are given as upper limits, the corresponding estimates of the X-ray-to-optical index are upper limits, too\footnote{The specific luminosities at 2 keV of the uncertain objects without X-ray counterparts have been computed using specific 2-8 keV flux upper limits, evaluated as a function of their position within the Chandra field of view \citep[see Figure 25 by][]{xue11}; we only considered the sources whose specific UV luminosities are computable.}.

\begin{figure}
   \centering
   \includegraphics[width=\hsize]{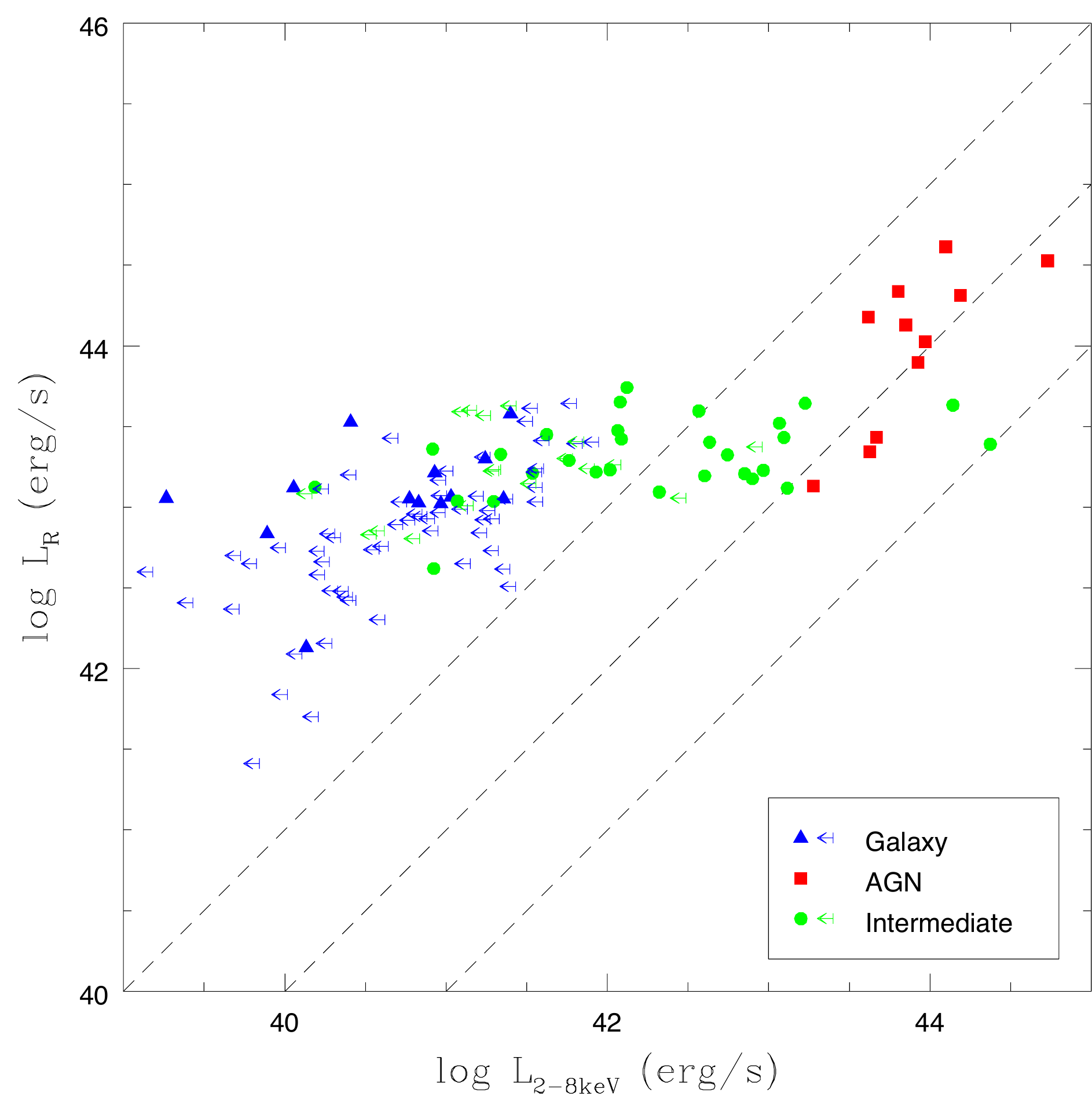}
\caption{$L_R$ versus $L_{2-8keV}$ diagram. Dashed lines indicate, from left to right, the ratios $\log (X/R)$$=$$-$1, $\log (X/R)$$=$0, and $\log (X/R)$$=$+1. Red, green, and blue symbols are defined as in Figure \ref{figalphaox}, and reported in the legend. All the intermediate source $L_R$ estimates refer to their total luminosities.}
\label{figXR}
\end{figure}

Figure \ref{figalphaox} shows the $\alpha_{ox}$ versus $\log L_{250nm}$ of AGN, galaxies, intermediate sources, and uncertain objects.  As a reference, we show the best-fit relations for AGN by \citet{just07} and \citet{luss10} (not corrected for the host galaxy contribution). We note that i) the distributions of AGN and galaxies are completely separated: the galaxy $\alpha_{ox}$ estimates are on average $\sim$one unit lower than the AGN reference relations, corresponding to X-ray luminosities $\sim$10$^2$-10$^3$ times lower than the average levels expected for active galactic nuclei; ii) most of the uncertain objects are low-luminosity sources and are close to the galaxy locus; iii) the intermediate sources are distributed on a wide region between the AGN and galaxy loci, partly overlapping them. This appears to be due to the different ratios between the AGN and host galaxy optical/UV continua of the sources. The host galaxy contamination on the average $\alpha_{ox}$-$\log L_{250nm}$ anti-correlation slope is discussed in \citet{vagn13}. We point out that the estimates of $\log L_{250nm}$ and $\alpha_{ox}$ correspond to the nuclear emission for the AGN and some of the intermediate objects (Figure \ref{figalphaox}), derived by the two-component fit of the SEDs, while are referred to the total UV luminosity for the other sources.

Also after the subtraction of the optical/UV galaxy contribution, part of the intermediate sources are still far from the AGN locus. This can be due, at least in part, to X-ray obscuration for type-2 sources. For a few sources it has been possible, on the basis of the available spectra from the XMM-CDFS Deep Survey, to compute de-absorbed X-ray luminosities and corresponding $\alpha_{ox}$ estimates, which are indicated in the plot by arrows from the absorbed to the corrected values. This correction gives X-ray-to-optical index values close to the reference $\alpha_{ox}$-$L_{250nm}$ relations. The intermediate sources can therefore be considered as AGN that are optically diluted by the host galaxy starlight and/or X-ray absorbed objects, and these effects can explain their low X-ray/UV ratio.

The distributions of the different classes of objects can also be shown in a $\log L_R$ versus $\log L_{2-8keV}$ plot (Figure \ref{figXR}). The R-band luminosities are derived from the EIS magnitudes, while the X-ray luminosities are computed from the Chandra 2-8 keV fluxes. As expected \citep[e.g.][]{alex02, xue11}, the AGN are concentrated in the $-$1<$\log (X/R)$<+1 region, and the galaxies present lower X-ray/optical ratios. The intermediate objects are distributed both in the $-$1<$\log (X/R)$<+1 stripe and outside --- $\log(X/R)$<$-$1, which is more typical of normal or star-forming galaxies --- as a consequence of the different level of optical dilution due to the host galaxy starlight. No attempt has been made to evaluate their R-band nuclear component. 

The possibility of selecting low-luminosity AGN candidates within the galaxy locus by means of optical variability analyses has been discussed by \citet{trev07b, trev08a}. After spectroscopic confirmation, such variable candidates are typically found to display ambiguous spectra between star-forming galaxies and low-luminosity AGN \citep[narrow emission line galaxies, NELG,][]{trev08b, bout09}. However, their variable character is an additional clue to their AGN nature. Several of our intermediate sources show ambiguous spectra as well, as referred to before \citep[e.g. LEX, ELG,][]{szok04, trei09}. A UV variability analysis for these sources would improve their characterisation. We carried out a preliminary analysis limited to those sources that have at least four measurements (including stacking determinations) in the UVW1 filter. Adopting a procedure similar to the one described in \citet{trev08a}, we estimate 12 variable candidates out of 453 objects in the magnitude interval 18.5$\leq$UVW1$\leq$23.5, corresponding to a 2.6\% fraction, which is comparable to the result by \citet{trev08a}, who selected 132 variable sources out of a reference sample of 7267 objects ($\approx$2\%). A more complete assessment of the variability will be done elsewhere.

\begin{figure}
   \centering
   \includegraphics[width=\hsize]{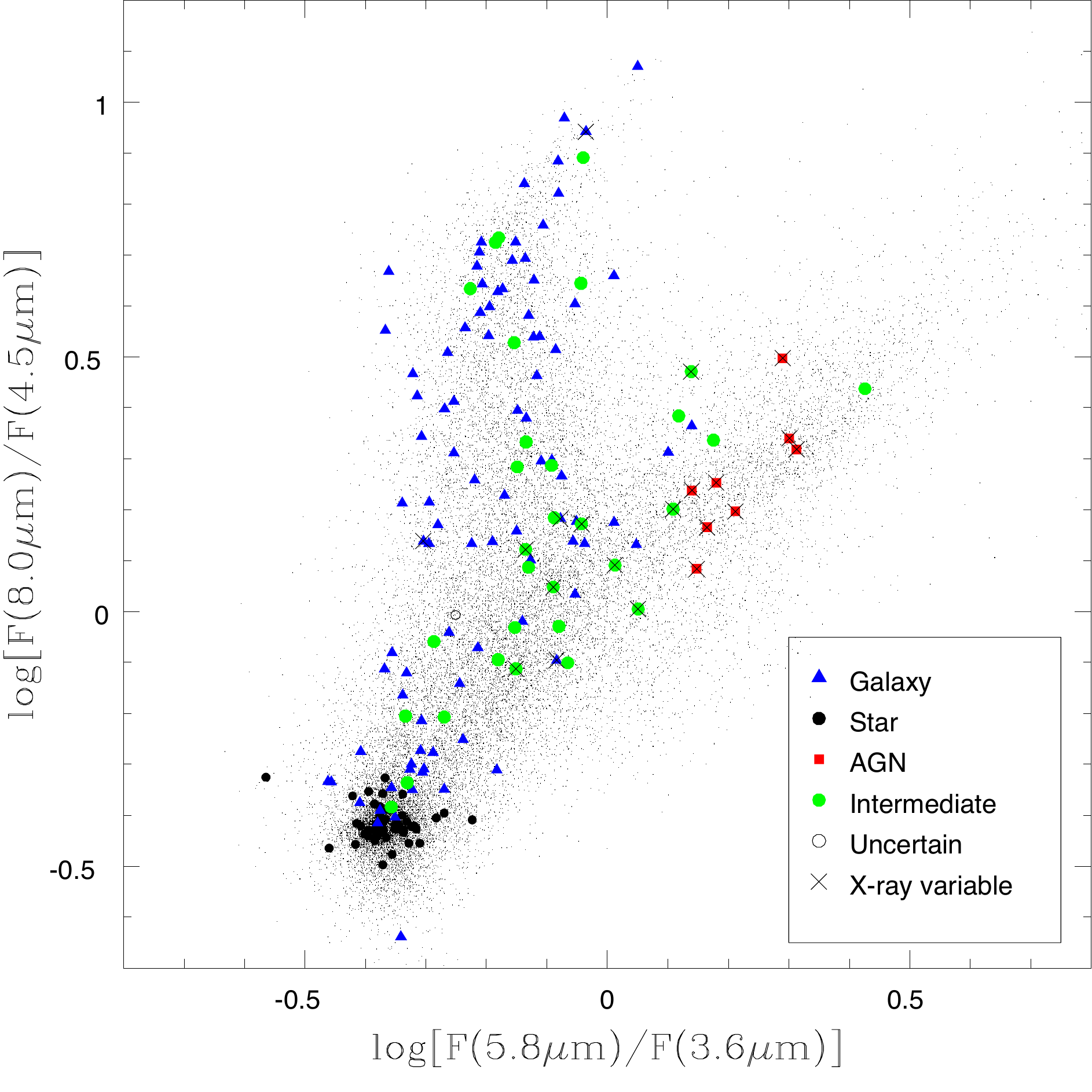}
\caption{The 8.0/4.5 $\mu$m flux ratio versus the 5.8/3.6 $\mu$m flux ratio. The small black dots refer to the SWIRE sample analysed by \citet[][and 2014 in preparation]{vacc10}. The largest symbols represent the XMMOMCDFS sources with available IR data from the previous sample, with the same meaning as the symbols in Figure \ref{figcolourUV}. In addition, the crosses indicate X-ray variable sources according to \citet[][and 2014 in preparation]{paol04}. Symbols are also reported in the legend.}
\label{figIR}
\end{figure}

To characterise the intermediate objects better and reinforce the indications of an AGN nature, it is also possible to use a diagnostic tool based on an IR colour-colour plot proposed by \citet{lacy04} and refined by \citet{donl12}, shown in Figure 12. The distribution of the XMMOMCDFS sources is compared to the data from the SWIRE \citep[][Vaccari et al. in preparation]{vacc10} survey. The reference diagram shows two branches, a nearly-vertical one mainly occupied by normal and star-forming galaxies, while most AGN are usually found in the inclined branch to the right. In fact, the QSOs in our catalogue are located in that region, but so are some of the intermediate sources. In the figure, we have also marked some sources that have been found to be X-ray variable according to the analysis by \citet{paol04} (and updated in Paolillo et al. 2014 in preparation): this constitutes an additional clue to their AGN nature, which we will investigate in more detail elsewhere. Moreover, we notice that the X-ray variability is also present for a few galaxies that possibly host low-luminosity AGN.

%

\subsection{Contents of the catalogue}\label{columns}

\begin{table*}
\caption{Column names in the XMMOMCDFS catalogue}             
\label{tabcolumncat}      
\centering                          
\begin{tabular}{r l r l r l}        
\hline\hline                 
Number & Name & Number & Name & Number & Name \\    
\hline                        
   1 & Cat\_srcnum & 34 & UVW1\_CR & 67 & V\_Dec \\      
   2 & Cat\_srcname & 35 & UVW1\_CR\_ERR & 68 & V\_POSERR \\
   3 & Tot\_UV\_RA & 36 &UVW1\_AB\_FLUX & 69 & V\_SIGNIF \\
   4 & Tot\_UV\_Dec & 37 & UVW1\_AB\_FLUX\_ERR & 70 & V\_CR  \\
   5 & UVW2\_OBS/STACK & 38 & UVW1\_AB\_MAG & 71 & V\_CR\_ERR \\
   6 & UVW2\_RA & 39 & UVW1\_AB\_MAG\_ERR & 72 & V\_AB\_FLUX \\
   7 & UVW2\_Dec & 40 & UVW1\_GroupSize & 73 & V\_AB\_FLUX\_ERR \\
   8 & UVW2\_POSERR & 41 & U\_OBS/STACK & 74 & V\_AB\_MAG \\
   9 & UVW2\_SIGNIF & 42 & U\_RA & 75 & V\_AB\_MAG\_ERR \\
   10 & UVW2\_CR & 43 & U\_Dec & 76 & V\_GroupSize \\
   11 & UVW2\_CR\_ERR & 44 & U\_POSERR & 77 & objID\_EIS \\
   12 & UVW2\_AB\_FLUX & 45 & U\_SIGNIF & 78 & EIS\_filter \\
   13 & UVW2\_AB\_FLUX\_ERR & 46 & U\_CR & 79 & objID\_COMBO \\
   14 & UVW2\_AB\_MAG & 47 & U\_CR\_ERR & 80 & Class\_COMBO \\
   15 & UVW2\_AB\_MAG\_ERR & 48 & U\_AB\_FLUX & 81 & CIF \\
   16 & UVW2\_GroupSize & 49 & U\_AB\_FLUX\_ERR & 82 & objID\_GALEX \\
   17 & UVM2\_OBS/STACK & 50 & U\_AB\_MAG & 83 & objID\_Chandra \\
   18 & UVM2\_RA & 51 & U\_AB\_MAG\_ERR & 84 & Full\_FLUX\_Chandra \\
   19 & UVM2\_Dec & 52 & U\_GroupSize & 85 & Soft\_FLUX\_Chandra \\
   20 & UVM2\_POSERR & 53 & B\_OBS/STACK & 86 & Hard\_FLUX\_Chandra \\
   21 & UVM2\_SIGNIF & 54 & B\_RA & 87 & Pind\_Chandra \\
   22 & UVM2\_CR & 55 & B\_Dec & 88 & Class\_Chandra \\
   23 & UVM2\_CR\_ERR & 56 & B\_POSERR & 89 & objID\_XMM \\
   24 & UVM2\_AB\_FLUX & 57 & B\_SIGNIF & 90 & FLUX\_XMM \\
   25 & UVM2\_AB\_FLUX\_ERR & 58 & B\_CR & 91 & z\_spect \\
   26 & UVM2\_AB\_MAG & 59 & B\_CR\_ERR & 92 & z\_spect\_ref \\
   27 & UVM2\_AB\_MAG\_ERR & 60 & B\_AB\_FLUX & 93 & logL250nm \\
   28 & UVM2\_GroupSize & 61 & B\_AB\_FLUX\_ERR & 94 & logL2keV \\
   29 & UVW1\_OBS/STACK & 62 & B\_AB\_MAG & 95 & X-ray-to-optical-index \\
   30 & UVW1\_RA & 63 & B\_AB\_MAG\_ERR & 96 & Adopted\_Class \\
   31 & UVW1\_Dec & 64 & B\_GroupSize & 97 & Notes \\
   32 & UVW1\_POSERR & 65 & V\_OBS/STACK & & \\
   33 & UVW1\_SIGNIF & 66 & V\_RA & & \\
\hline                                   
\end{tabular}
\end{table*}

In this section, we briefly present the columns that compose the UV main catalogue of the XMM-CDFS Deep Survey (Table 7), and the supplementary catalogue concerning the alternative identifications for CIF=3 sources (Table 8). The two tables are available in electronic form at the CDS. The columns of the main catalogue are listed in Table \ref{tabcolumncat}, while the supplementary catalogue contains a subset of them, specified at the end of the section.

Column 1 is a reference sequential source number, arbitrarily assigned to the object (Cat\_srcnum). Column 2 contains the name of the source (Cat\_srcname). Columns 3-4 are the right ascension and declination of the object in decimal degrees, computed by averaging the mean coordinates of the detections in each UV band (Tot\_UV\_RA, Tot\_UV\_Dec). Columns 5-76 provide the average astrometry, detection significance, and photometry of the measurements in each filter ({\it f}); in particular, {\it f}\_OBS/STACK states whether the source has been detected on individual observations (o) or stacked images (s); {\it f}\_RA and {\it f}\_Dec refer to the average coordinates of the detections in decimal degrees; {\it f}\_POSERR is the 1-$\sigma$ mean position uncertainty in arcsecs; {\it f}\_SIGNIF gives the average significance of the detections; {\it f}\_CR and {\it f}\_CR\_ERR represent the mean count rate and the corresponding uncertainty; {\it f}\_AB\_FLUX and {\it f}\_AB\_FLUX\_ERR are the mean flux density (corresponding to the effective wavelength of the {\it f} band) and its uncertainty; {\it f}\_AB\_MAG and {\it f}\_AB\_MAG\_ERR give the average AB magnitude and its uncertainty; {\it f}\_GroupSize refers to the number of individual observations (or stacked images) in which the source has been detected (without bad quality flags).

Columns 77-78 refer to the name of the EIS cross-identification in the shortest wavelength available band (objID\_EIS) and the name of the corresponding filter (EIS\_filter).

Columns 79-80 give the sequential number of the COMBO-17 cross-identification (objID\_COMBO) and the corresponding photometric classification\footnote{G: Galaxy; S: Star; WD: White Dwarf; Q: AGN; G/U: Galaxy (Uncl!); G/S: Galaxy (Star?); Q/G: QSO (Gal?); S.O.: Strange Object.} (Class\_COMBO).

CIF (column 81) is defined in Section \ref{vali} and refers to the reliability of the identification within the EIS and the COMBO-17 survey\footnote{The values 1, 3, and 4 indicates the sources with no unique cross-identification, analogously to the notes related to the cross-correlations with the other CDF-S surveys (Column 97).}.

Columns 82 gives the GALEX source-number (objID\_GALEX).

Columns 83-88 refer to the Chandra 4 Ms source catalogue counterpart: objID\_Chandra is the source identification; Full\_FLUX\_Chandra, Soft\_FLUX\_Chandra, and Hard\_FLUX\_Chandra are the observed frame fluxes\footnote{Negative values indicate upper limits of the corresponding fluxes.} in the 0.5-8 keV, 0.5-2 keV, and 2-8 keV bands, in units of erg/s/cm$^2$; Pind\_Chandra is the effective photon index; Class\_Chandra is the combined X-ray/optical classification of the source\footnote{G: Galaxy; S: Star; A: AGN.} (see Section \ref{combclas}).

Columns 89-90 are the entry numbers of the XMM-Newton 2-10 keV point-source catalogue counterpart (objID\_XMM) and the 2-10 keV flux (FLUX\_XMM) in units of erg/s/cm$^2$.

Columns 91 gives a spectroscopic estimate of redshift (z\_spect) obtained from the Arizona CDFS Environment Survey or the ESO CDF-S Master Catalogue; column 92 (z\_spect\_ref) flags the reference from which the redshift estimate is derived\footnote{A: ACES \citep{coop12}. For the ESO CDF-S Master Catalogue we refer to the specific surveys: L \citep{lefe05}, M \citep{mign05}, P \citep{pope09, bale10}, Si \citep{silv10}, Sz \citep{szok04}, V \citep{vanz08}.}.

Columns 93-95 are the log of the specific luminosity at 250 nm (logL250nm), the log of the specific luminosity at 2 keV (logL2keV), and the $\alpha_{ox}$ estimate (X-ray-to-optical index) of the UV source with X-ray counterpart from the Chandra 4 Ms source catalogue. Columns 93 and 95 refer to the total emission of the source, without extracting the nuclear component.

Column 96 gives the combined classification\footnote{G: Galaxy; S: Star; WD: White Dwarf; Q: AGN; I: Intermediate; U: Uncertain/unclassified.} (Adopted\_Class).

Finally, ambiguous COMBO-17, GALEX, Chandra, XMM-Newton, or ACES cross-identifications are noted in column 97 (Notes). The sequential number of the alternative, discarded identification is reported in parentheses. We notice that ACES adopts the same sequential numbers as COMBO-17.

Concerning the supplementary catalogue (Table 8), its first two columns are the sequential number and name of the sources already described for the main catalogue, but both include an additional letter that indicates the alternative XMM-OM identification(s) associated with the same EIS/COMBO-17 source (CIF=3). None of these alternative identifications have UVW2 measurements; however, the corresponding columns are also reported in the supplementary catalogue to maintain the same numbering. This catalogue contains 44 alternative identifications for 38 EIS/COMBO-17 sources in total and is limited to the columns 1-76, with the same meaning as in the main catalogue.

%

\section{Discussion}\label{concl}

The XMMOMCDFS catalogue was compiled to provide complementary UV photometry of known sources in the Chandra Deep Field-South. These observations have the advantage of a better spatial resolution compared to GALEX and a wider field of view with respect to HST. The comparison of XMM-OM with GALEX and EIS magnitudes in similar bands confirms the validity of the XMM-OM calibration. The XMM-OM data allow a better characterisation of the SED of the sources in the UV domain, at wavelengths between the GALEX bands and the visible region, also complementing multi-wavelength studies of the SEDs from the IR to the X-ray bands.

The catalogue contributes to the detection of fainter sources than those already present in the archival source lists corresponding to the individual observations, through the stacking of exposure images from groups of consecutive observations. The additional sources detected through the stacking are $\sim$31\% of the total number in each UV band, corresponding to an increase of $\sim$45\% with respect to the sources detected and validated within the individual observations.

The catalogue includes 1129 sources, 1031 of which compose the subset of reliable cross-identifications (``good subsample''): $\sim$79\% of the sources in the good subsample are galaxies, while $\sim$6\% are classified as AGN or intermediate sources. The latter are optically classified as galaxies by the COMBO-17 survey, while showing AGN-like X-ray properties. The analysis of their optical/UV SED allowed us to distinguish galactic and nuclear components for many of these sources. They appear to be AGN that are optically diluted by the dominant host galaxy starlight. Moreover, the available optical spectra provided by the spectroscopic surveys show ambiguous characteristics, i.e. narrow and weak emission lines, which could be attributed to low-luminosity AGN or star formation activity. Part of the intermediate sources are also to be considered as obscured AGN, and for a few of them it has been possible to evaluate their corrected nuclear X-ray fluxes, so estimating standard X-ray/UV ratios. Further confirmation of the AGN nature of these objects and/or the selection of variable candidates among the galaxy population may be obtained through a UV variability analysis, which could provide additional evidence of AGN activity for some of the sources and a characterisation of the properties of the UV variability itself through the use of the structure function. A preliminary analysis limited to the UVW1 band yields 12 variable candidates. In a future work, we will more completely analyse the UV (and optical) variability of our sources on the basis of the 33 available XMM-OM observations. This will allow us to select AGN candidates by means of their variability, analogously to previous \citep{trev08a, geza13} and ongoing \citep{deci14,falo14} works. In particular, the efficiency of variability observations has been discussed for selecting AGN candidates with extended images (as could be for many of our intermediate sources), which would not be selected by colour or other criteria \citep{bers98, trev08a}. The time sampling of the light curves of many of the catalogued sources is well suited to investigating their possible variable character. We will complement this with other variability information already present in the CDF-S from optical \citep{trev08a}, UV \citep{geza13}, or X-ray \citep{youn12} observations. The variability of the sources will also be examined in terms of the X-ray/UV ratio in a companion paper devoted to the structure function analysis of the X-ray-to-optical index for a subsample of AGN and intermediate sources.

%

\begin{acknowledgements}

We acknowledge funding from PRIN/MIUR-2010 award 2010NHBSBE. We acknowledge support from the Italian Space Agency under the contract ASI/INAF/009/10/0/1 and by the Italian National Institute for Astrophysics (INAF) through PRIN-INAF 2011 (“Black hole growth and AGN feedback through the cosmic time”). M.A. acknowledges the hospitality by the European Space Astronomy Centre during Autumn 2013, while part of this work was being done. We thank Luciana Bianchi for useful advise and discussion about the GALEX data. We also thank the referee for valuable remarks and comments. This research has made use of data and/or software provided by the High Energy Astrophysics Science Archive Research Center (HEASARC), which is a service of the Astrophysics Science Division at NASA/GSFC and the High Energy Astrophysics Division of the Smithsonian Astrophysical Observatory.

\end{acknowledgements}

%

\bibliographystyle{aa}
\bibliography{bib}{}

\end{document}